\documentclass[bibyear]{aa}
\usepackage{graphicx}
\usepackage{txfonts}
\usepackage{hyperref}
\usepackage{mathabx}
\usepackage{natbib}
\bibpunct{(}{)}{;}{a}{}{,} 
\usepackage{tabularx}
\usepackage{multirow}
\usepackage{multicol, blindtext}
\usepackage{color}
\usepackage{amsmath}
\usepackage{xcolor}
\usepackage[toc,title,page]{appendix} 
\usepackage{geometry}
\usepackage{arydshln}
\usepackage{amssymb}
\usepackage{ulem}

\newcommand{\BJDTDB}{\mathrm{BJD}_\mathrm{TDB}}

\newcommand{\Porb}{P_{\mathrm{orb}}}
\newcommand{\Prots}{P_\mathrm{rot,\star}}
\newcommand{\Protp}{P_\mathrm{rot,p}}
\newcommand{\rp}{r_\mathrm{p}}
\newcommand{\mpl}{m_\mathrm{p}}

\newcommand{\Ms}{M_\mathrm{\star}}
\newcommand{\esin}{\mathrm{\sqrt{e}\cdot \sin \omega_0}}
\newcommand{\ecos}{\sqrt{e}\cdot\cos \omega_0}
\newcommand{\Loves}{k_\mathrm{{2,\star}}}
\newcommand{\Lovep}{k_\mathrm{{2,p}}}

\begin{document}

   \title{Characterizing WASP-43b's interior structure: unveiling tidal decay and apsidal motion}

   \author{L. M. Bernab\`o \inst{\ref{inst:PF@DLR}}$^{\href{https://orcid.org/0000-0002-8035-1032}{\includegraphics[scale=0.5]{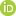}}}$
          \and Sz. Csizmadia \inst{\ref{inst:PF@DLR},\ref{inst:ELKH}}$^{\href{https://orcid.org/0000-0001-6803-9698}{\includegraphics[scale=0.5]{Images/orcid.jpg}}}$
          \and A.~M.~S.~Smith\inst{\ref{inst:PF@DLR}}$^{\href{https://orcid.org/0000-0002-2386-4341}{\includegraphics[scale=0.5]{Images/orcid.jpg}}}$
          \and J.-V. Harre\inst{\ref{inst:PF@DLR}}$^{\href{https://orcid.org/0000-0001-8935-2472}{\includegraphics[scale=0.5]{Images/orcid.jpg}}}$ 
          \and Sz. K\'alm\'an \inst{\ref{inst:Konkoly}, \ref{inst:HUN}, \ref{inst:ELTE}} $^{\href{https://orcid.org/0000-0003-3754-7889}{\includegraphics[scale=0.5]{Images/orcid.jpg}}}$  
          \and J. Cabrera\inst{\ref{inst:PF@DLR}}$^{\href{https://orcid.org/0000-0001-6653-5487}{\includegraphics[scale=0.5]{Images/orcid.jpg}}}$ 
          \and H. Rauer \inst{\ref{inst:PF@DLR},\ref{inst:FU}}$^{\href{https://orcid.org/0000-0002-6510-1828}{\includegraphics[scale=0.5]{Images/orcid.jpg}}}$
          \and D. Gandolfi \inst{\ref{inst:Torino}}$^{\href{https://orcid.org/0000-0001-8627-9628}{\includegraphics[scale=0.5]{Images/orcid.jpg}}}$
          \and L. Pino \inst{\ref{inst:INAF_Fi}}
          \and D. Ehrenreich \inst{\ref{inst:Geneve1},\ref{inst:Geneve2}} $^{\href{https://orcid.org/0000-0001-9704-5405}{\includegraphics[scale=0.5]{Images/orcid.jpg}}}$ 
           \and A. Hatzes \inst{\ref{inst:TLS}}$^{\href{https://orcid.org/0000-0002-3404-8358}{\includegraphics[scale=0.5]{Images/orcid.jpg}}}$
          }

   \institute{
   \label{inst:PF@DLR} Institute of Planetary Research, German Aerospace Center (DLR), Rutherfordstrasse 2, 12489 Berlin, Germany
   \and \label{inst:ELKH} ELKH-SZTE Stellar Astrophysics Research Group, H-6500 Baja, Szegedi \'ut Kt. 766, Hungary
   \and \label{inst:Konkoly} Konkoly Observatory, Research Centre for Astronomy and Earth Sciences, HUN-REN, MTA Centre of Excellence, Konkoly-Thege Miklós út 15–17., 1121, Hungary
    \and \label{inst:HUN} HUN-REN-ELTE Exoplanet Research Group, Szombathely, Szent Imre h. u. 112., 9700, Hungary
    \and \label{inst:ELTE} ELTE Eötvös Loránd University, Doctoral School of Physics, Budapest, Pázmány Péter sétány 1/A, 1117, Hungary
    \and \label{inst:FU} Institut f{\"u}r Geologische Wissenschaften, Freie Universit{\"a}t Berlin, 12249 Berlin, Germany
    \and \label{inst:Torino} Dipartimento di Fisica, Universit{\`a} degli Studi di Torino, Via Pietro Giuria, 1, 10125 Torino, Italy
   \and \label{inst:INAF_Fi} INAF – Osservatorio Astrofisico di Arcetri, Largo Enrico Fermi 5, 50125 Firenze, Italy
   \and \label{inst:Geneve1} Observatoire Astronomique de l’Université de Genève, Chemin
Pegasi 51, 1290 Versoix, Switzerland
\and \label{inst:Geneve2} Centre Vie dans l’Univers, Faculté des sciences, Université de
Genève, Quai Ernest-Ansermet 30, 1211 Genève 4, Switzerland
 \and \label{inst:TLS}  Th{\"u}ringer Landessternwarte Tautenburg, Sternwarte 5, 07778 Tautenburg, Germany\\
   \email{lia.bernabo@dlr.de, liamarta.bernabo@gmail.com}\\
   }

 \date{Submitted 26 August 2024; accepted 3 January 2025}

 \abstract
   {Recent developments in exoplanetary research highlight the importance of Love numbers in understanding their internal dynamics, formation, migration history and their potential habitability. Love numbers represent crucial parameters that gauge how exoplanets respond to external forces such as tidal interactions and rotational effects. By measuring these responses, we can gain insights into the internal structure, composition, and density distribution of exoplanets. The rate of apsidal precession of a planetary orbit is directly linked to the second-order fluid Love number, thus we can gain valuable insights into the mass distribution of the planet.}
   {In this context, we aim to re-determine the orbital parameters of WASP-43b - in particular, orbital period, eccentricity, and argument of the periastron - and its orbital evolution. We study the outcomes of the tidal interaction with the host star: whether tidal decay and periastron precession are occurring in the system.}
   {We observed the system with HARPS, whose data we present for the first time, and we also analyse the newly acquired JWST full-phase light curve. We fit jointly archival and new radial velocity and transit and occultation mid-times, including tidal decay, periastron precession and long-term acceleration in the system.}
   {We detected a tidal decay rate of $\dot{P}_a$= (-1.99$\pm$0.50) 
   ms yr$^{-1}$ and a periastron precession rate of $\dot{\omega}$=
   0.1727$^{+0.0083}_{-0.0089}$)$^\circ$ d$^{-1}$ = (621.72 $^{+29.88}_{-32.04}$)$^{\prime\prime}$d$^{-1}$. This is the first time that both periastron precession and tidal decay are simultaneously detected in an exoplanetary system. The observed tidal interactions can neither be explained by the tidal contribution to apsidal motion of a non-aligned stellar or planetary rotation axis nor by assuming non-synchronous rotation for the planet, and a value for the planetary Love number cannot be derived. Moreover, we exclude the presence of a second body - e.g. a distant companion star or a yet undiscovered planet -, down to a planetary mass of $\gtrsim$0.3 $M_J$ and up to an orbital period of $\lesssim$ 3\,700 days. We leave the question of the cause of the observed apsidal motion open.}
   {}

   \keywords{exoplanets - interior structure - tidal interaction - periastron precession - tidal decay}
   
  \titlerunning{WASP-43b orbital evolution}
  \authorrunning{Bernab\'o et~al.}
   \maketitle

\section{Introduction}
Among the abundance of exoplanets confirmed thus far, hot Jupiters - characterized by their high mass, size and proximity to their host stars - inhabit the most extreme environments. These characteristics generate large transit and radial velocity signals, contributing to the precise determination of various planetary system parameters, including planetary mass, radius, surface gravity, orbital distance, and stellar density. The identification of such planetary characteristics has significantly advanced our theoretical comprehension of these systems, which may deviate from the architectural norm observed in our Solar System, allowing for the exploration of planetary composition and internal structure, offering valuable insights into the formation, evolution, and migration history of these planets. Tidal forces play a crucial role in this context, influencing the variety and composition of these systems (see e.g. \citealp{Valencia2006,Sotin2007,Hatzes2015}). Moreover, the investigation of tidal interactions between a star and a hot Jupiter unveils insights into the internal structures of celestial bodies. Despite these observational advantages, like the large signal-to-noise ratio, the origin of hot Jupiters is still unclear (see for example \citealp{Dawson2018}).\\
\indent The knowledge of the mass and radius of an exoplanet - and, consequently, its bulk density - plays a crucial role in categorizing it, e.g. as a rocky planet or a hot gaseous Jupiter-like body. However, this information alone is insufficient to unveil their internal structure, encompassing layers, thickness, and composition (see e.g. \citealp{Seager2007} and reference therein). Potential solutions often exhibit high degeneracy, presenting multiple interior compositions that differ qualitatively but equally match the observed data. 
In the context of Jupiter-like close-in planets, Love numbers (\citealp{Love1911}) emerge as valuable information (see e.g. \citealp{Becker2013}). Specifically, the second-order fluid Love number $\Lovep$ is directly proportional to the mass concentration towards the centre of the celestial body. \cite{Padovan2018} and \cite{Baumeister2020} demonstrated that incorporating this parameter as an input in the interior structure model significantly diminishes the ambiguity associated with potential internal structures.\\

\noindent In this context, WASP-18Ab (\citealp{Csizmadia2019}) and WASP-19Ab (\citealp{Bernabo2024}) were chosen as case studies. We now include WASP-43b in our investigation. WASP-43b is an Ultra-hot Jupiter (UHJ; with period P< 1 d) with $\simeq$ 2 $M_\mathrm{\Jupiter}$ orbiting around a K7V-type star 87 parsecs away, with a period of $\simeq$ 0.813 d.\\
\indent While orbital decay has been confirmed in WASP-12b (see~\citealp{Maciejewski2016},~\citealp{Yee2020} and reference therein), the case of WASP-43b is still under discussion. 
A detailed description of the literature on WASP-43b's (non-)detections of tidal decay is given in Section \ref{sec:model_results}. WASP-43b is one of the best candidates for observing decay and, in general, the outcomes of tidal interaction between the planet and the host star due to its high planet-to-star mass ratio and proximity to the host star.\\
\indent This paper follows the approach and method of~\cite{Bernabo2024} and we aim to investigate the orbital evolution of WASP-43b and the planetary Love number. We collected archival transit and occultation data and performed a new radial velocity (RV) study of the system with the High Accuracy Radial velocity Planet Searcher (HARPS). We also analyse for the first time four RV datasets which are public but have never been published yet.\\

\noindent The paper is organized as follows. In Section~\ref{sec:data} we present transit and occultation timing observations, previous RV studies as well as new HARPS RV data. In Section~\ref{sec:model_results} we fit the RV and transit and occultation mid-times with a model including a long-term acceleration of the system, tidal decay and apsidal motion in the planetary orbit. In Section \ref{sec:discussion} we discuss the implications and possible explanations of our findings, such as the non-synchronous rotation of WASP-43b, the presence of a second putative planetary body in the system and the stellar rotation axis inclination. 

\section{Data}
\label{sec:data}
\subsection{Transits and Occultations}
\label{subsec:TrOcc}
The mid-transit times used in this paper are listed in Table~\ref{Tab:transits1} and the mid-occultation times in Table~\ref{Tab:occultations}. These data are taken from previously published literature and from the {\it Exoplanet Transit Database} (ETD) \footnote{\url{http://var2.astro.cz/ETD/}} (\citealt{Poddany2010}) and {\it ExoClock}\footnote{\url{https://www.exoclock.space}} (\citealp{Kokori2022}) database. In these two databases, we discarded data with incomplete transits or low-quality (according to the quality flag assigned by the databases) observations. In case any transit event from such databases was not published yet, the observer was contacted directly to gain permission to use the data. We also analysed a full phase curve taken by the James Webb Space Telescope (JWST; \citealp{Gardner2006}) on the 1$^\mathrm{st}$ and 2$^\mathrm{nd}$ December 2022 (published in \citealp{Bell2024}) with the Transit Light Curve Modeler (TLCM, \citealp{Csizmadia2020}) and unpublished transits from Transiting Exoplanet Survey Satellite (TESS; \citealp{TESS2016}) Sector 62. More details are given in Sections \ref{sec:JWST} and \ref{sec:TESS62}. \\
\indent For each transit and occultation mid-time, the time unit was checked and, if necessary, transformed into~$\BJDTDB$ (Barycentric Julian Date in the Barycentric Dynamical Time) with the help of the online conversion tool developed by~\cite{Eastman2010}\footnote{\url{https://astroutils.astronomy.osu.edu/time/}}.

\subsubsection{JWST phase curve}
\label{sec:JWST}
\begin{figure}
    \centering
    \includegraphics[width=0.5\textwidth]{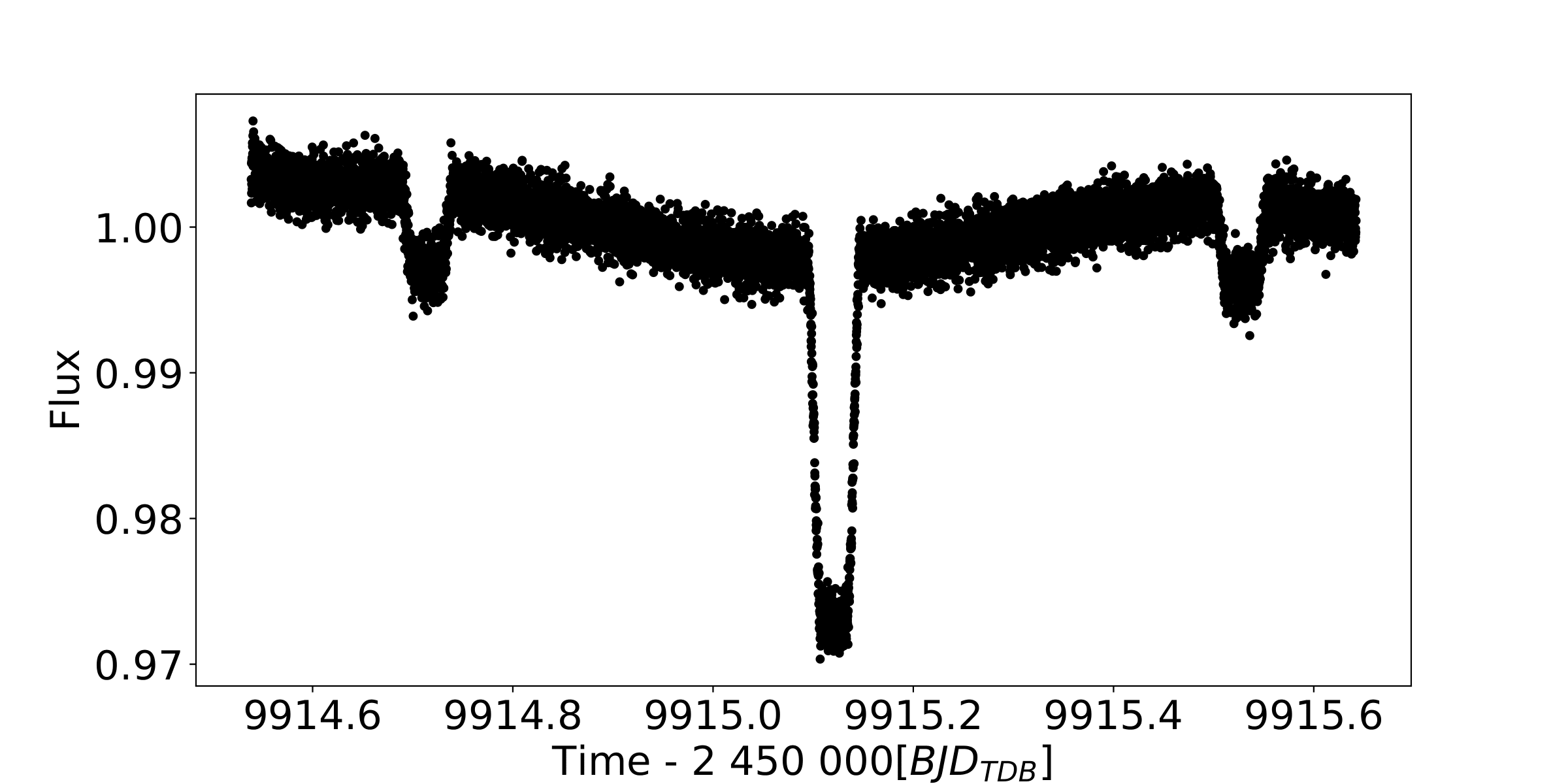}
     \includegraphics[width=0.5\textwidth]{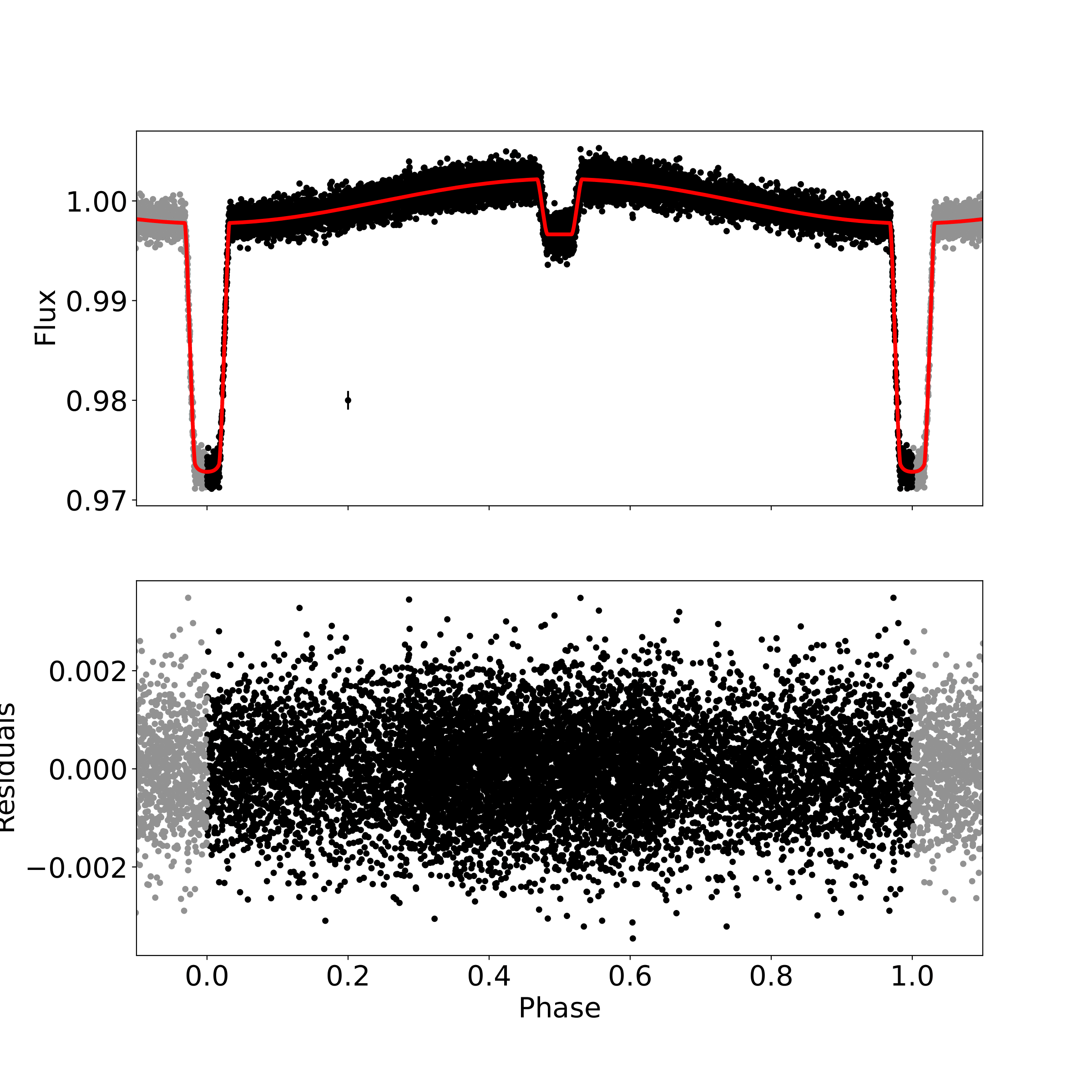}
    
    \caption{JWST transit and occultations.\\
    {\it Top plot} JWST full-phase observations in time.\\
    {\it Middle plot} Phase-folded data points in black and best fit in red. To keep the figure clean, errorbars are not plotted, but a typical errorbar value is shown at location (0.2,0.98). Grey points are the same as black points, but shifted by $\pm$1.0 in phase to make the full-phase clearer. \\ 
    {\it Low plot} Residuals of the fit, having subtracted the red noise, with a standard deviation of $\simeq$0.00097.}
    \label{fig:JWST}
\end{figure}

We fitted the full-phase light curve acquired by JWST with the MIRI instrument between 6.5 and 7.0 $\mu$m. The data were published by \cite{Bell2024}, but no study has performed a comprehensive analysis of transit, occultation and phase curve yet. We used TLCM to retrieve the transit and occultation parameters (\citealp{Csizmadia2020}). We used limb darkening priors from \texttt{ExoCTK}\footnote{\url{https://exoctk.stsci.edu/limb_darkening}} and we fitted the reflection effect (including a shift with respect to the sub-stellar point), red noise, white noise, the intensity ratio, the beaming effect and the ellipsoidal effect. The correlated noise is handled with a wavelet-based method (\citealp{CarterWinn2009,Csizmadia2023}). The full phase curve is modelled with a lambertian fit.\\
\indent First, we fitted the full phase curve and then the transit and two occultations separately. The transit and occultation mid-times which resulted from the individual fit of the three events are also reported in Table \ref{Tab:transits1} and \ref{Tab:occultations} and are used in the fit in Section \ref{sec:model_results}. In particular, the two occultations are essential since they come after $\simeq$8 years of no reported mid-occultation time and fundamentally help to discern the Transit Timing Variation (TTV) trend between tidal decay and apsidal motion (see Section \ref{sec:complete_model}, Figure \ref{fig:Tiddecay_apsmotion} and its relative discussion for more details).\\
\indent In Table \ref{tab:JWST}, we show the value of the fitting parameters and in Figure \ref{fig:JWST} we show the best fitting of the full-phase and its residuals. The transit duration is derived as  1.2226$^{+0.0064}_{-0.0070}$ hours. 
\begin{table*}[]
    \centering
    \caption{Fitting parameters of the JWST light curve.}
    \begin{tabular}{l c c c}
    \hline \hline
        Parameter & Symbol [Units] & Value & Priors \\
        \hline
        Mid-transit time & $\mathrm{T_{tr}}$ [$\BJDTDB$-2~450~000] & 9915.12076 $\pm$ 0.00032 & $\mathcal{U}$ [9914.0;9916.0] \\
        Sidereal period & $\mathrm{P_s}$ [d] & 0.81346$^{+0.00031}_{-0.00030}$ & $\mathcal{U}$ [0.8125; 0.8144]\\
        Scaled semi-major axis & $\mathrm{a/R_\star}$ & 4.834 $^{+ 0.050}_{-0.045}$ & $\mathcal{U}$ [0; 10] \\
        Planet-to-star radius ratio & $\mathrm{r_p/R_\star}$ &  0.1568$^{+0.0012}_{-0.0010}$ & $\mathcal{U}$ [0.5; 0.5] \\
        Planet-to-star intensity ratio & $\mathrm{I_{p}/I_\star}$ & 0.0445 $^{+0.0099}_{-0.0113}$& $\mathcal{U}$ [0; 1] \\
        Impact parameter & b & 0.6676$^{+0.0082}_{-0.0111}$ & $\mathcal{U}$ [0; 1]\\
        & $\mathrm{\sqrt{e}\cdot \sin \omega}$ & 0.0065$^{+0.0440}_{-0.0559}$ & $\mathcal{U}$ [-1; 1] \\
        & $\mathrm{\sqrt{e} \cdot \cos \omega}$ & -0.0082$^{+0.0073}_{-0.0098}$ & $\mathcal{U}$ [-1; 1] \\
        White noise & $\mathrm{\sigma_w}$ & 0.0007722$^{+0.0000047}_{-0.0000049}$ & $\mathcal{U}$ [0; 1] \\
        Red noise & $\mathrm{\sigma_r}$ & 0.02044 $\pm$ 0.00053 & $\mathcal{U}$ [0; 1]  \\
        Limb darkening & u$_+$ & 0.168 $\pm$ 0.024& \\
        & u$_-$ & -0.05 $\pm$ 0.14 & \\
        \hline
    \end{tabular}
    \tablefoot{Median and 1$\sigma$ errorbars of the fitting parameters of the full-phase JWST lightcurve between 6.5 and 7.0 $\mu$m. Limb darkening parameters $u_{+}$ and $u_{-}$ are not fitting parameters, but derived from the limb darkening parameters $A$ and $B$ (see \citealp{Kalman2024}), so their priors are not reported in the Table.}
    \label{tab:JWST}
\end{table*}

\subsubsection{TESS Sector 62}
\label{sec:TESS62} 
\begin{figure}
    \centering
    \includegraphics[width=0.5\textwidth]{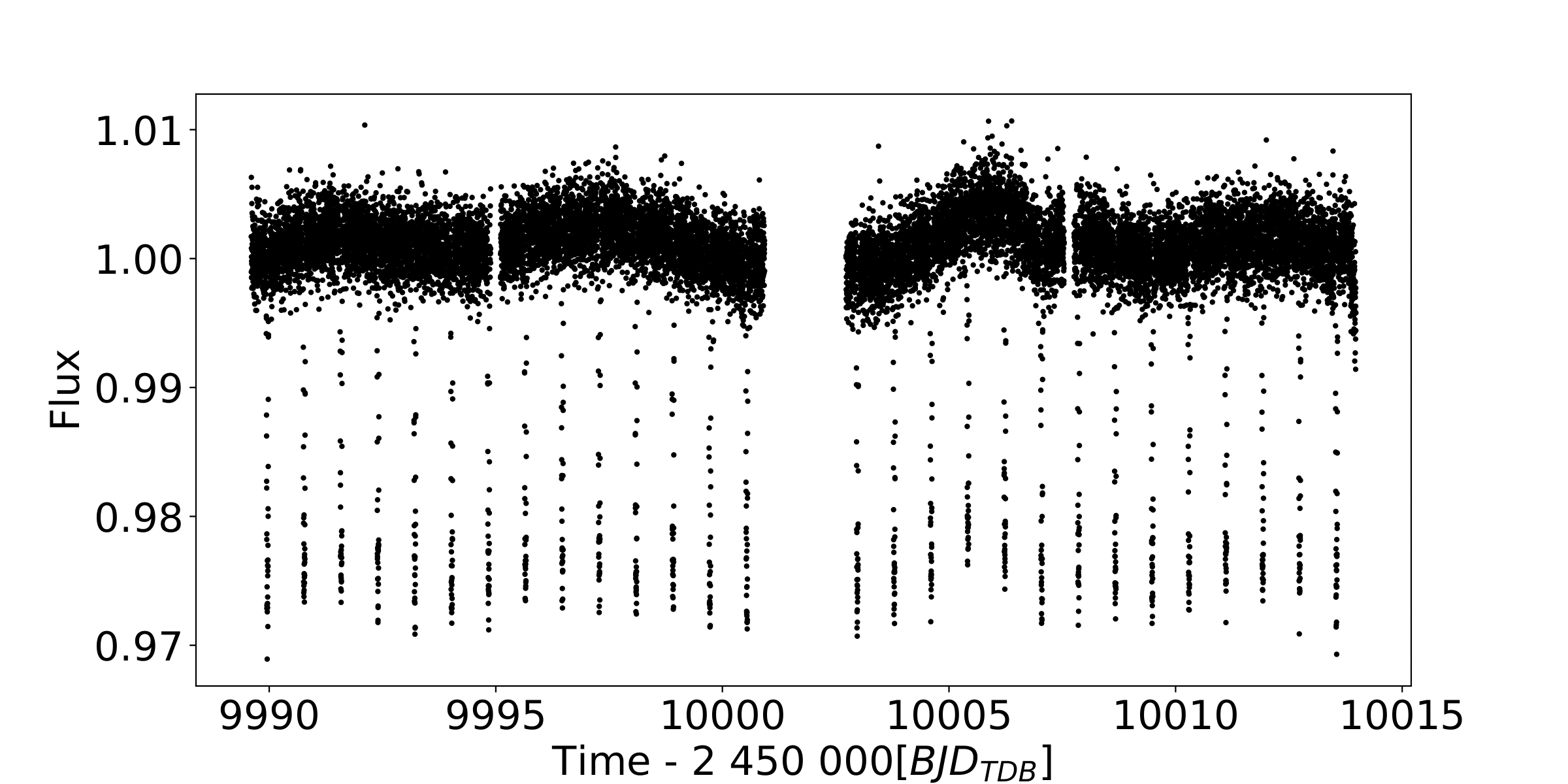}
     \includegraphics[width=0.5\textwidth]{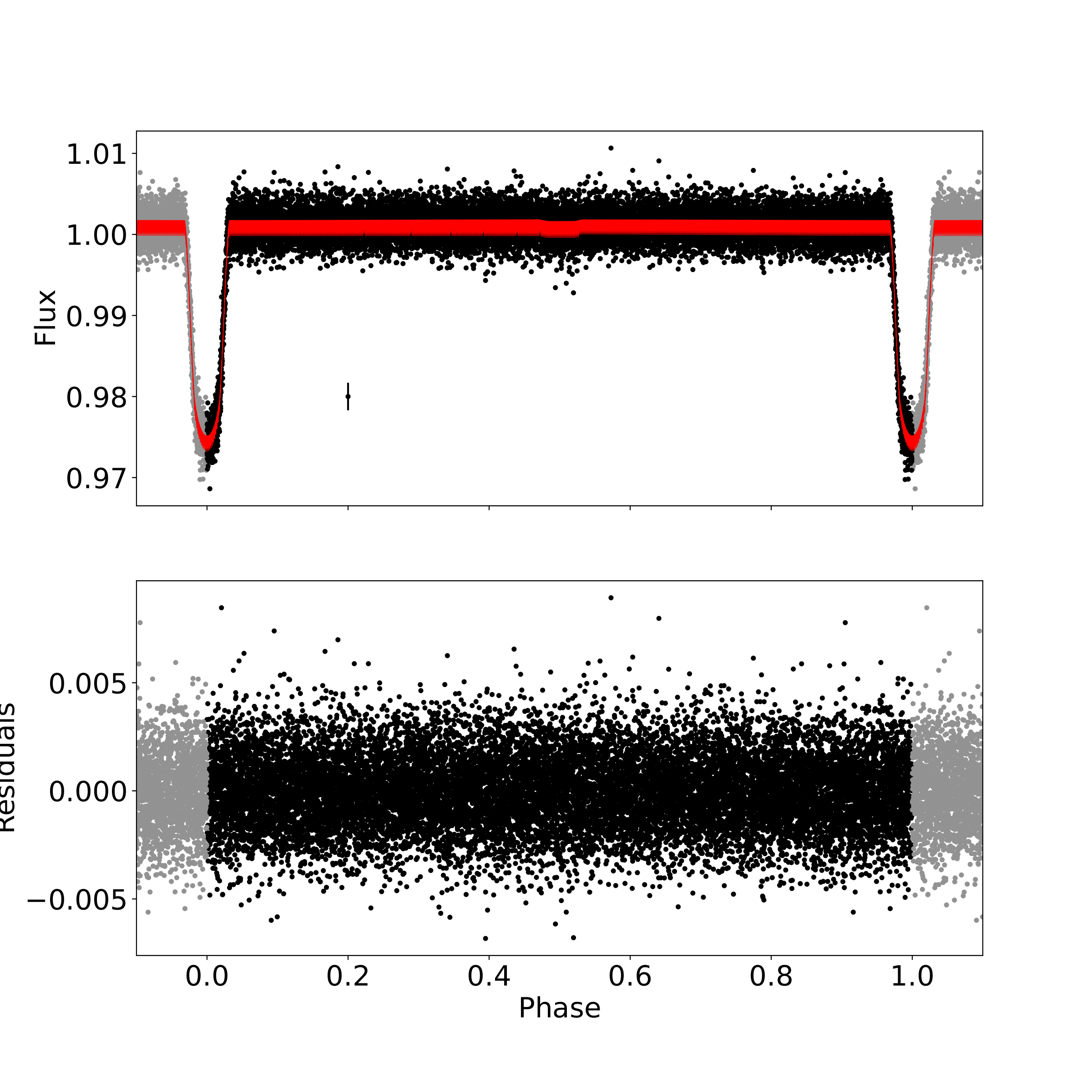}
    
    \caption{TESS Sector 62.\\
    {\it Top plot} TESS 120-seconds full-phase observations in time.\\
    {\it Middle plot} Phase-folded data points in black and best fit in red. The typical errorbar value is shown at location (0.2,0.98). Grey points are shifted by $\pm$1.0 in phase. The occultation is slightly visible at phase $\simeq$0.5.\\ 
    {\it Low plot} Residuals of the fit, having subtracted the red noise, with a standard deviation of $\simeq$0.0017.}
    \label{fig:TESS62}
\end{figure}

We also included previously unpublished observations of WASP-43b made by TESS. Sector 62 was observed in February and March 2023 at 120-second and 20-second cadence. 28 transits of WASP-43b were captured with both exposure times. We analysed them with TLCM two mid-transit times were obtained, corresponding with the two exposure times. The fitting parameters are reported in Table \ref{tab:TESS62}, while in Figure \ref{fig:TESS62} we show the 120s phase-folded transits and their residuals.
\begin{table*}[]
\caption{Fitting parameters of TESS light curves.}
    \centering
    \begin{tabular}{l c c c c}
    \hline \hline
        Parameter & Symbol [Units] & Value & Value & Priors \\
        & & 20 s & 120 s & \\
        \hline
        Mid-transit time & $\mathrm{T_{tr}}$ [$\BJDTDB$-2~450~000] & 9989.96040$^{+0.00011}_{-0.00017}$ & 9989.960418$^{+0.000090}_{-0.000091}$ & $\mathcal{U}$ [9980.0; 9982.0] \\
        Sidereal period & $\mathrm{P_s}$ [d] & 0.8134745 $\pm$ 0.0000052 & 0.8134744$\pm$0.0000050 & $\mathcal{U}$ [0.8125; 0.8144]\\
        Scaled semi-major axis & $\mathrm{a/R_\star}$ & 4.853$^{+0.054}_{-0.057}$ & 4.851$^{+0.051}_{-0.052}$ & $\mathcal{U}$ [0; 10] \\
        Planet-to-star radius ratio & $\mathrm{r_p/R_\star}$ & 0.15773$^{+0.00063}_{-0.00061}$ & 0.15777$^{+0.00059}_{-0.00060}$ & $\mathcal{U}$ [0.5; 0.5] \\
        Planet-to-star intensity ratio & $\mathrm{I_{p}/I_\star}$ & 0.0046$^{+0.0038}_{-0.0031}$ & 0.0048$^{+0.0041}_{-0.0030}$ & $\mathcal{U}$ [0; 1] \\
        Impact parameter & b & 0.6543$^{+0.0089}_{-0.0097}$& 0.6545$^{+0.0083}_{-0.0093}$ & $\mathcal{U}$ [0; 1]\\
        White noise & $\mathrm{\sigma_w}$ & 0.001741$^{+0.000011}_{-0.000010}$ & 0.001742$\pm$0.000010 & $\mathcal{U}$ [0; 1] \\
        Red noise & $\mathrm{\sigma_r}$ & 0.03190$\pm$0.00099 & 0.03187$^{+0.00099}_{-0.00098}$ & $\mathcal{U}$ [0; 1]  \\
         Limb darkening & u$_+$ & 0.661$\pm$0.030 & 0.663$\pm$0.031 & \\
        & u$_-$ & 0.31$\pm$0.013 & 0.32$\pm$0.13& \\
        \hline
    \end{tabular}
    \tablefoot{Median and 1$\sigma$ errorbars of the fitting parameters of TESS Sector 62 for both exposure times. As for JWST full-phase, limb darkening parameters $u_{+}$ and $u_{-}$ are not fitting parameters, but derived from the limb darkening parameters $A$ and $B$, so their priors are not reported in the Table.}
    \label{tab:TESS62}
\end{table*}

\subsection{Radial Velocities}
\label{subsec:RVs}
In this work, we used literature and yet unpublished RV data, as well as newly acquired HARPS data. In Table~\ref{Tab:RV_data} we list details on the datasets, as labelled in the table, while in Table~\ref{tab:RVs1} we report the seven Radial Velocity datasets, where all times are corrected to $\BJDTDB$. In Table \ref{Tab:RV_data} we highlight the number of out-of-transit as the datapoints we use since our model does not fit the Rossiter-McLaughlin effect. In Section \ref{sec:RM} we separately fit such an effect to constrain the stellar spin-orbit angle.\\
\begin{table}[]
    \centering
    \caption{RV datasets.}
    \begin{tabular}{c c l l}
      ID & $N_\mathrm{obs}$ & Instrument & Reference  \\
        & (out of transit) & & \\
        \hline 
        1 & 15 (14) & CORALIE &~\cite{Hellier2011} \\
        2 & 8 (7) & CORALIE &~\cite{Gillon2012} \\ 
        3 & 67 (21) & HARPS-S & This work (PI: Triaud)\\
        4 & 40 (19) & HARPS-S &~\cite{Esposito2017} \\
        5 & 21 (10) & HARPS-N & This work \\
        6 & 56 (36) & HARPS-S & This work (PI: Ehrenreich)\\
        7 & 128 (48) & ESPRESSO & This work (PI: Pino)\\
        8 & 19 (18) & HARPS-S & This work (PI: Csizmadia)\\
        \hline
    \end{tabular}
    \tablefoot{Details on the radial velocity datasets used for our analysis: ID number (as in this work, in chronological order), the total number of observations and, in parenthesis, those out of transit, instrument and reference.}
    \label{Tab:RV_data}
\end{table}

\noindent The datasets are labelled as:
\begin{enumerate}
\item[ID 1]~\cite{Hellier2011}, discovery paper: 15 radial-velocity measurements between January and July 2010 as a follow-up with the CORALIE spectrograph at the Swiss 1.2-metre Leonhard Euler Telescope at ESO’s La Silla Observatory;
\item[ID 2]~\cite{Gillon2012}: 8 spectra acquired in February and March 2011 by CORALIE; 
\item[ID 3]~This work: 67 unpublished spectra taken in 2012 by the High Accuracy Radial velocity Planet Searcher (HARPS-S, \citealp{Mayor2003}) to measure the spin-orbit angle via the Rossiter-McLaughlin effect and orbital eccentricity. PI: A. Triaud, program ID: 089.C-0151;
\item[ID 4]~\cite{Esposito2017}: 32 spectra acquired by HARPS-S during and around one transit in 2013 and 8 additional spectra later in 2013 and 2015. 27 of these datapoints were also published in~\cite{Bonomo2017}; 
\item[ID 5]~This work: 21 unpublished HARPS-N (\citealp{Cosentino2012}) spectra, acquired during and around one transit in 2015. Program ID: OPT15B$\textunderscore$19;
\item[ID 6]~This work: unpublished spectra taken in 2016 by HARPS-S to resolve the planetary atmosphere with transit spectroscopy. The total number of available RV spectra is actually 101, however, we selected only the first and third night of observations, due to the low signal-to-noise ratio of the second night data. PI: D. Ehrenreich, program ID: 096.C-0331;
\item[ID 7]~This work: 128 unpublished ESPRESSO spectra taken in 2020 to study the escape mechanism of WASP-43b's atmosphere. PI: L. Pino, program ID: 0102.C-0820;
\item[ID 8]~This work: 19 newly acquired RV points were taken for this study using HARPS-S. PI: Sz. Csizmadia, program ID: 0104.C-0849. 
\end{enumerate}

\begin{figure} 
    \centering
    \includegraphics[width=0.5\textwidth]{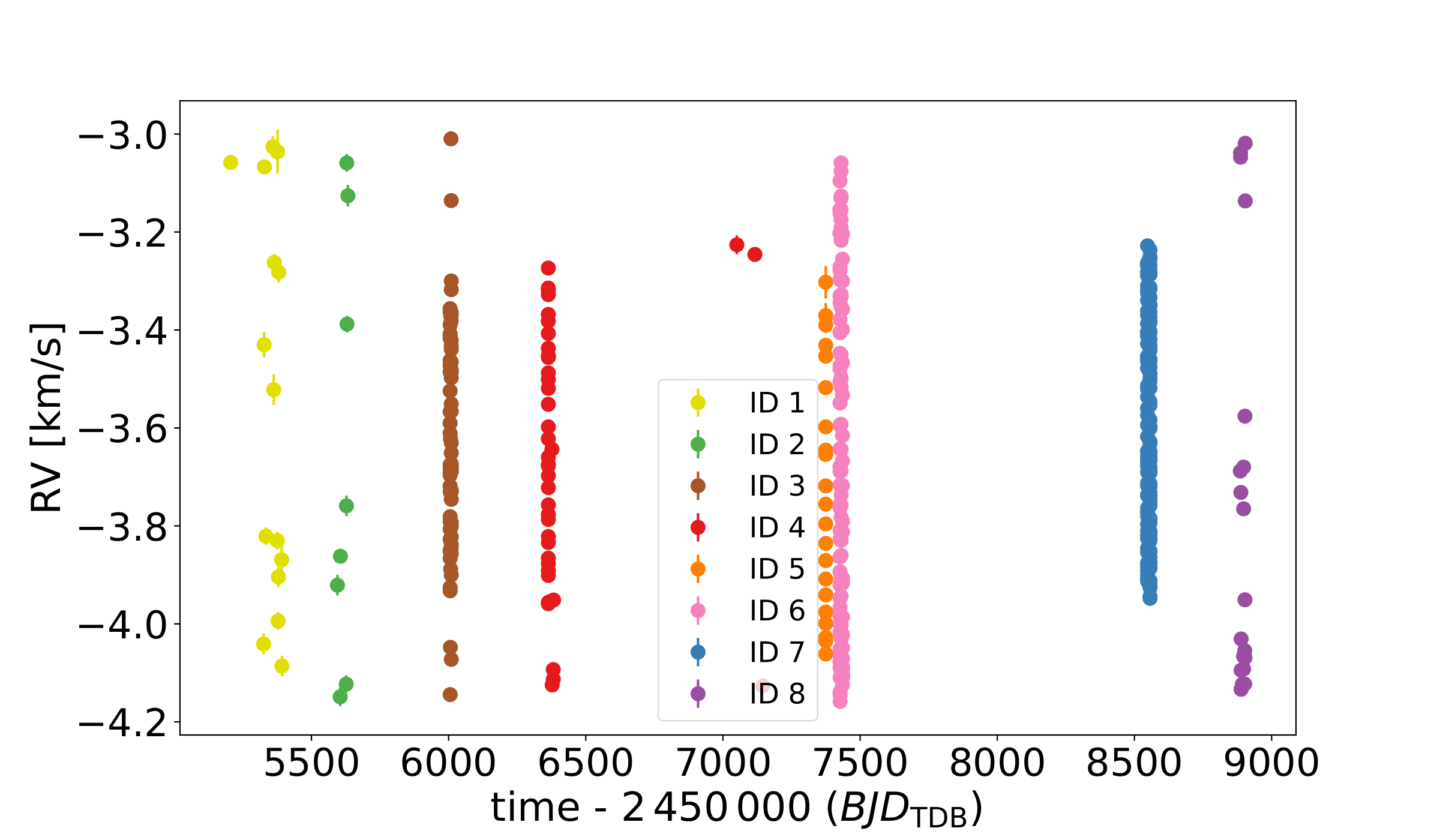}
    \caption{Previsously published and new radial velocity observations of WASP-43. The different data sets are plotted with different colours showing the time distribution of the RV observations.}
    \label{fig:RV_data}
\end{figure}

These RV observations cover a time span of ten years and are plotted in Figure~\ref{fig:RV_data}, where the offsets among them were subtracted for a clear plot.\\ 

\section{Data analysis}
\label{sec:model_results}

\subsection{Eccentricity study}
\label{sec:eccentricity}
Most close-in planets are found to be on a circular orbit due to the strong tidal interactions with the host star leading to the quick circularization of the orbit (see e.g \citealp{Zahn2008}). In the case of WASP-18Ab (\citealp{Csizmadia2019}) and WASP-19Ab (\citealp{Bernabo2024}), a long-distance companion star is supposed to be a possible cause of the slightly eccentric orbit (see also Appendix B in \citealp{Bernabo2024}). A requirement for apsidal motion to take place is that the orbital eccentricity is not zero, therefore we analyse the orbital eccentricity as a precondition for our study. Other USP planets are likely in a slightly eccentric orbit: apart from the two aforementioned cases, 55 Cnc e with $e$=0.05$\pm$0.03, GJ 367b with $e$=0.06$^{+0.07}_{-0.04}$ (\citealp{Goffo2023}) and TOI-500b with $e$=0.063$^{+0.068}_{-0.044}$ (\citealp{Serrano2022}), LHS 1678b 0.033$^{+0.035}_{-0.023}$ (\citealp{Silverstein2024}), which are all are in a multi-planetary system. \\
\indent In many studies on WASP-43b, its orbit is assumed to be circular (see for example \citealp{Hellier2011,Chen2014}). Through transit, occultation and RV fitting, \cite{Hellier2011} gave an upper limit of 0.04 at 3$\sigma$ for the eccentricity and adopted a circular orbit for the fit, \cite{Gillon2012} deduced $e$=0.0035$^{+0.0060}_{-0.0025}$, and assumed the result consistent with a fully circularized orbit. \cite{Blecic2014} found a mid-occultation phase of 0.5001$\pm$0.0004 and $e$=0.010$^{+0.010}_{-0.007}$ and \cite{Chen2014} an offset occultation phase of 0.4907$^{+0.0036}_{-0.0027}$. \\
\indent We can now use the most complete and extended RV, transit and occultation dataset up to date, including unpublished data, to determine the orbital eccentricity of WASP-43b. \\



\begin{figure}
    \centering
    \includegraphics[width=0.5\textwidth]{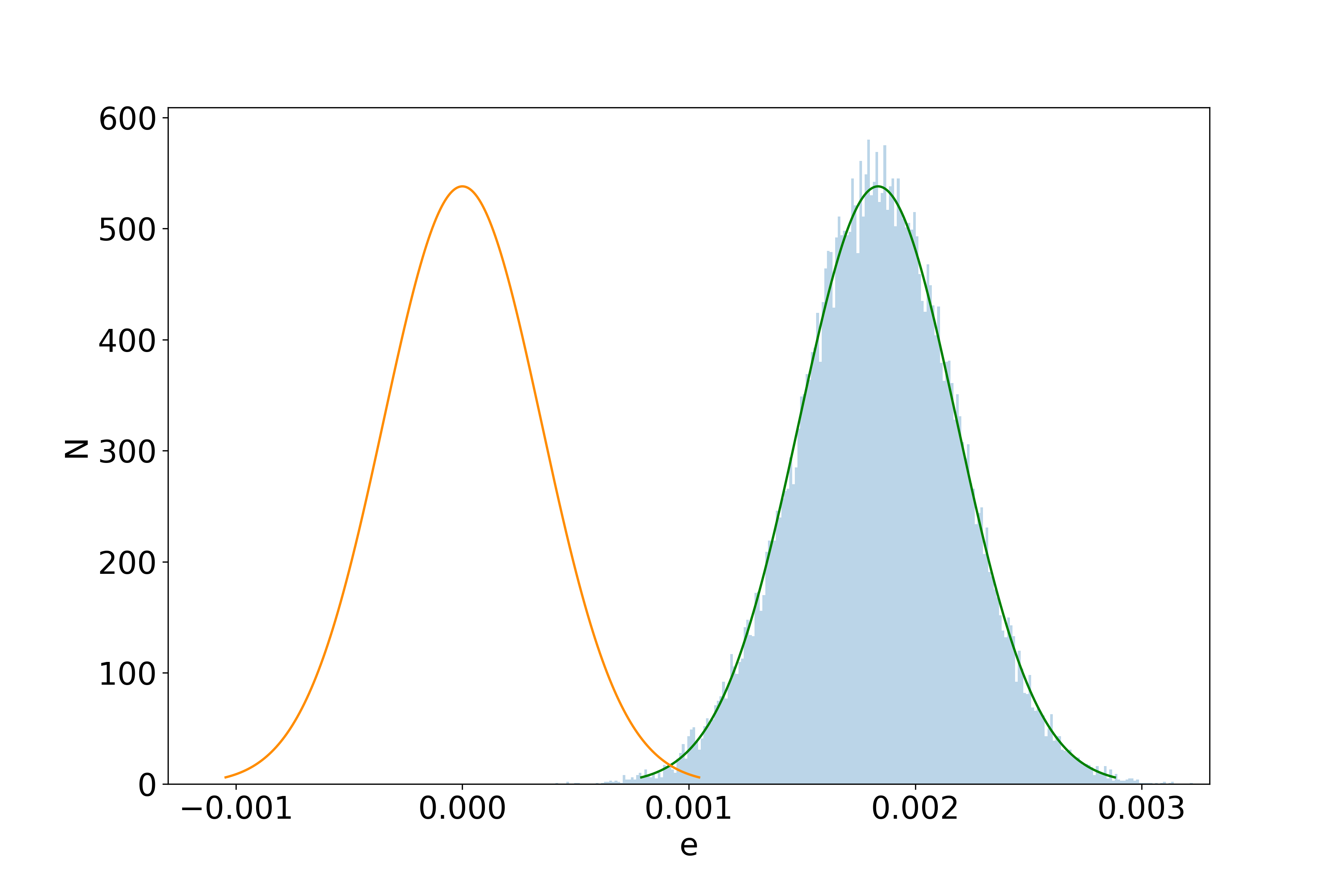}
    \caption{Eccentricity significance. The posterior distribution of the eccentricity is shown in blue in transparency. The normal distribution with $\mu$=0.00188 and $\sigma$=0.00035 in blue is compared to a normal distribution with the same variance, shifted to have its mean at 0 (orange curve). The spikes in the posterior distributions are highly investigated regions, such as local minima, where the chains were probably stuck before converging to the global minimum.}
    \label{fig:eccentricity}
\end{figure}

\noindent We fitted the complete RV, transit and occultation mid-times dataset with a one planet circular and an eccentric model and compared the results. Notably, our analysis incorporates a significantly larger dataset than previous studies, utilizing 471 data points, the largest dataset compared to the aforementioned studies. When fitting an eccentric planet, the eccentricity resulted to be significantly different from zero: $e$=0.00290 $^{+0.00148}_{-0.00068}$, (1$\sigma$ errorbars), with the 5$\sigma$ uncertainty range extending to $^{+0.0092}_{-0.0025}$. We applied a Student's {\it t}-test to compare the posterior distribution with a null hypothesis of circular orbit. We obtained {\it p-value}$\sim 10^{-5}$ and {\it t-statistic} value of $\simeq$ 437. The cut-off for the {\it p-value} is typically set at $\alpha$=0.05. A lower {\it p-value} indicates a significant difference between the two distributions, meaning that the observed data provide strong evidence that the sample distribution is different from zero.\\
\indent We also calculated the Bayesian Information Criterion (BIC; \citealp{Schwarz1978}) of the two fits as:
\begin{equation}
    BIC = \chi^2 + k \ln (n)
\end{equation}
where $k$ is the number of fitting parameters and $n$ the number of datapoints that are fitted. Since the BIC disfavours the more complex models when $n$ is big, we also calculated the Akaike Information Criterion (AIC; \citealp{Akaike1998}):
\begin{equation}
    AIC = \chi^2 + 2 k
\end{equation}
With a difference of two fitting parameters ($\esin$ and $\ecos$, where $\omega_0$ is the periastron angle at epoch zero, namely here $T_0$=5528.86823 $\BJDTDB$), the BIC and AIC differences between the two fittings are both $\gtrsim$ 200, meaning that there is strong evidence that the eccentric orbit is preferred over the circular one for WASP-43b.\\
\indent Moreover, when fitting an eccentric orbit with tidal decay and periastron precession scenario (see Section \ref{sec:complete_model}), we obtained a better constraint on the eccentricity, as shown in Table \ref{tab:model_parameters}: $e$=0.00183 $\pm$ 0.00035, consistent with the previous result. This finding is significantly non-zero even at 5$\sigma$ confidence. Figure \ref{fig:eccentricity} shows the non-tailed normal posterior distribution of the eccentricity in blue and the same curve with the same standard deviation shifted to have its median at 0. The two Gaussians are clearly different. We also applied the student's{\it t}-test. The $p-value$ we obtained is <10$^{-5}$ and its $t-statistic$ value is >2000, meaning that the sample mean deviates from the mean of the null hypothesis by more than 2000 standard deviations and that the observed data are extremely unlikely under the null hypothesis.\\
\indent We then proceeded to the modelling of our data, under the assumption that an eccentric orbit can be fitted, and therefore apsidal motion can take place.

\subsection{One-planet model}
\label{sec:simple_models}
We modelled together RV data, transit and occultation mid-times like in~\cite{Csizmadia2019} and~\cite{Bernabo2024} with an $\texttt{idl}$ script composed of a Genetic Algorithm minimization which is followed by a Differential Evolution Markov-chain Monte Carlo (DE-MCMC,~\citealp{TerBraak2006}). The initial population of the Genetic Algorithm is composed of 200 individuals. The number of steps of the DE-MCMC procedure is 10$^5$ for each of the 10 chains. The burn-in phase discards the first 8$\cdot$10$^4$ steps of such a procedure and the posterior distributions and Gelman - Rubin test are made with the last 2$\cdot$10$^4$ iterations. \\

\noindent First of all, we investigated the simplest models to describe our joint dataset: the one-planet circular and eccentric fits, as partly described in the previous Section. We began by fitting a single planet on a circular orbit to the data, which provided an initial baseline model. This approach assumes that the planet’s orbit is perfectly circular, simplifying the mathematical representation and reducing the number of parameters. The fitting parameters are the RV semi-amplitude $K$ and the offset $V_\gamma$, as well as the instrumental offsets $V_{instr}$ for each RV dataset and the orbital period $P$, while the first transit epoch $T_0$ is assumed as in \cite{Hellier2011}: $T_0$=5528.86823 $\BJDTDB$. Without the inclusion of photometric and RV jitters, with 471 data points and 15 fitting parameters, we obtained a BIC value of $\simeq$2190 and an AIC of $\simeq$2132. However, the circular fit may not fully capture the dynamics of the system and tidal interactions that can take place. In the one-planet eccentric fit, two extra fitting parameters were included: $\esin$ and $\ecos$, as described in the previous Section. For this fit, the BIC value is $\simeq$1855 and AIC$\simeq$1789. 

\subsection{Tidal decay model}
\label{sec:tidaldecay}
\begin{figure}
    \centering
    \includegraphics[width=0.49\textwidth]{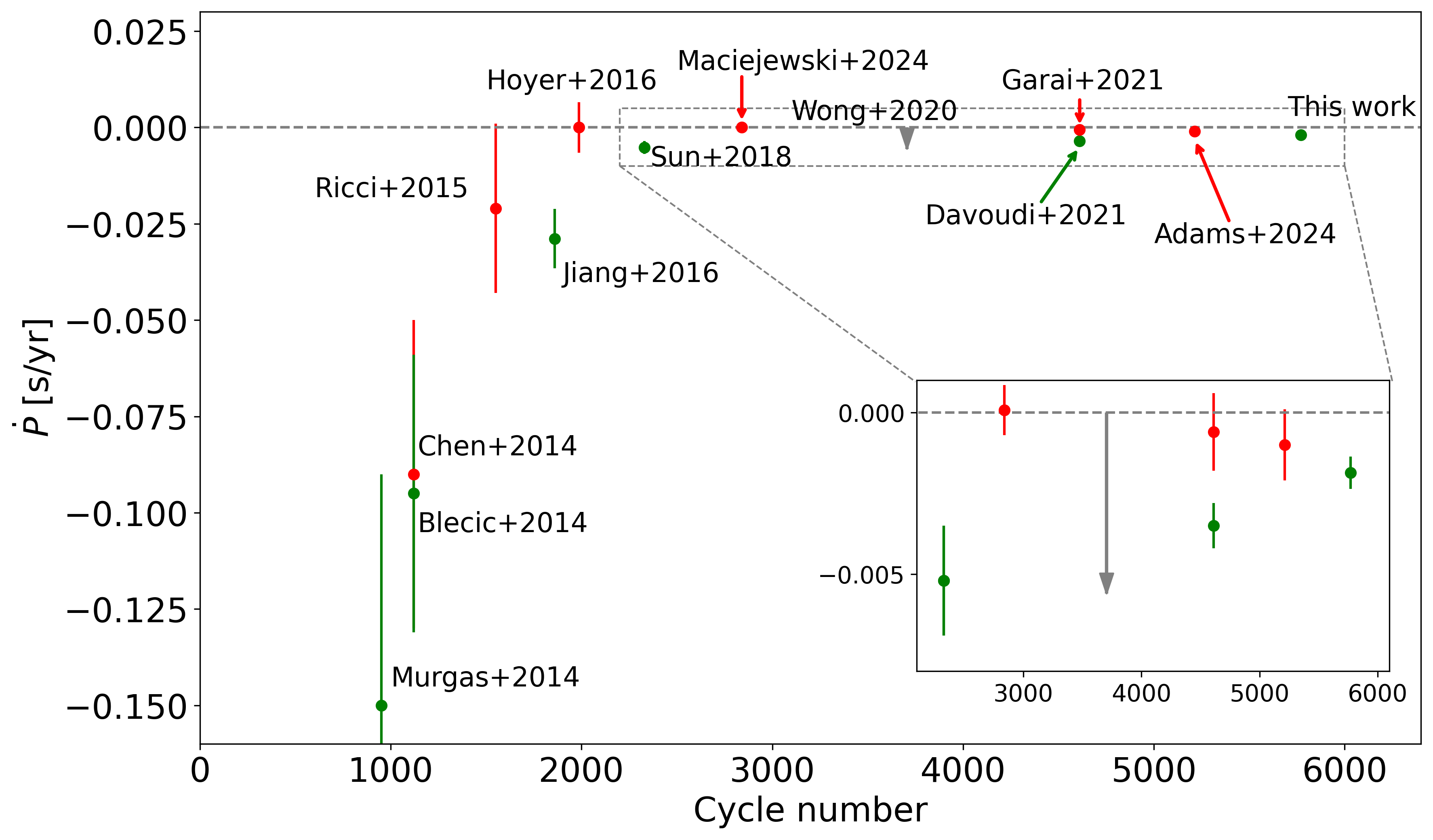}
    \caption{Literature retrievals of tidal decay rate and our results plotted as a function of cycle number, starting from \cite{Hellier2011} as epoch 0. Non-detections are shown in red, detections in green and the only upper limit in grey. The region between cycle numbers 2000 and 6000 is zoomed to show the small errorbars of the latest studies. Our result is shown in green because, even if BIC and AIC disfavour the tidal decay fit in comparison to the one-planet eccentric fit, in Section \ref{sec:complete_model} we detect the same tidal decay rate and such a fit is the preferred one. The errorbar on our value is particularly small because we fit TTVs and RVs simultaneously and because we make use of the longest dataset available up to date.
    }
    \label{fig:tid_decay_literature}
\end{figure}

 A typical effect that can appear in systems showing a strong tidal interaction between the host star and the planet, such as in WASP-12 (\citealp{Yee2020} and reference therein), is tidal decay of the planetary orbit. Tidal decay has been investigated in WASP-43b via transit timing variations. Some authors claimed the detection of tidal decay in the system, while other studies discarded it or provided an upper limit. Table \ref{tab:tidal_decay} summarizes the findings of the previous studies. Among these studies, we point out that \cite{Chen2014} found a decay rate of -0.09 $\pm$ 0.04 s yr$^{-1}$, however the BIC criterion does not prefer the quadratic fit over the linear fit. \cite{Ricci2015} obtained a decay rate of -0.021 $\pm$ 0.022 s yr$^{-1}$, but the small difference in $\chi^2$ between the linear and the quadratic fits of the Transit Timing Variations (TTVs) shows that statistically the latter brings no significant improvement. \\
\begin{table*}[]
    \centering
    \caption{Tidal decay results in literature.}
    \begin{tabular}{l c l}
    \hline \hline
        {\bf $\dot{\mathrm{P}}$} [s yr$^{-1}$] & {\bf Detection} [Y/N/u] & {\bf Reference} \\
        \hline
        -0.09 $\pm$ 0.04 & N & \cite{Chen2014} \\
        -0.15 $\pm$ 0.06 & Y & \cite{Murgas2014} \\
        -0.095 $\pm$ 0.036 & Y & \cite{Blecic2014} \\
        -0.021 $\pm$ 0.022 & N & \cite{Ricci2015}\\ 
        -0.0289 $\pm$ 0.0077 & Y & \cite{Jiang2016} \\
        -0.00002 $\pm$ 0.0066 & N & \cite{Hoyer2016}\\
        -0.0052 $\pm$ 0.0017 & Y & \cite{Sun2018} \\
        <-0.0056 & u & \cite{Wong2020} \\  
        -0.0035 $\pm$ 0.0007 & Y & \cite{Davoudi2021} \\
        -0.0006 $\pm$ 0.0012 & N & \cite{Garai2021}\\
        -0.0010 $\pm$ 0.0011 & N & \cite{Adams2024}\\
        (0.77 $\pm$ 7.76)$\cdot10^{-4}$ & N & \cite{Maciejewski2024}\\
         \hline
    \end{tabular}
    \tablefoot{Summaries of the previous results on the tidal decay scenario in the WASP-43 system, also plotted in Figure \ref{fig:tid_decay_literature}. The second column indicates if the studied detected [Y] or discarded [N] tidal decay in WASP-43b's orbit or gave an upper limit [u]. All results in the literature are obtained using TTVs only.}
    \label{tab:tidal_decay}
\end{table*}
\indent Figure \ref{fig:tid_decay_literature} shows the literature values for the tidal decay rate and our result. The cycle number is taken as the last transit event considered by the corresponding paper. Note the decreasing decay rate (and relative errorbars with time). \\

\noindent We investigated the tidal decay scenario by fitting RVs and TTVs simultaneously for the first time and using the longest dataset available ($\simeq$10 years). We implemented the script considering that the anomalistic period $P_a$ is a linearly time-varying parameter, with a time derivative $\dot{P}_a$: $P_a(t)$=$P_{a,0} + \dot{P}_a (t-T_0)$, where $T_0$ represents the first transit epoch and is not a fitting parameter, but fixed to $T_0$=5528.86823 $\BJDTDB$ (\citealp{Hellier2011}). $\dot{P}_a$ is linked to the change in the RV semi-amplitude $\dot{K}$ and through Kepler's Third Law to the time-derivative of the semi-major axis $\dot{a}$: 
\begin{equation}
    \frac{\dot{a}}{a(t)} = \frac{3}{2}\frac{\dot{P}_a}{P_a(t)}.
    \label{eq:ThirdLaw}
\end{equation}
\indent The RV curve has now the addition of the semi-amplitude $K(t)$ as a time-varying parameter, where its time-derivative $\dot{K}$ is derived from the combination of Equation \ref{eq:ThirdLaw} and the well-known definition of $K$:
\begin{equation}
    \dot{K}= - \frac{a \hspace{0.1cm} q \hspace{0.1cm} \sin i}{\sqrt{1-e^2}} \frac{\dot{P}_a}{P_a^2} = - K \frac{\dot{P}_a}{P_a},
    \label{eq:Kdot}
\end{equation}
where $q$ the planet-to-star mass ratio and $i$ the orbital inclination. \\
\indent TTVs are fitted with a quadratic trend, as in Equation 4 and 5 of \cite{Harre2023a} or as the third term of Equation \ref{eq:TTVs}.\\
\indent We obtained a tidal decay rate of (-1.85 $\pm$ 0.51) ms yr$^{-1}$, compatible with the value of Section \ref{sec:complete_model}. With 17 fitting parameters, the BIC value is $\simeq$1946 and AIC is $\simeq$1876.\\

\subsection{Long-term acceleration model}
\label{sec:longterm}
The next step in our analysis involved extending the model to consider the possibility of additional bodies in the system (see also Section \ref{sec:perturber}). In this case, the RV model included an extra long-term acceleration $\dot{V}_\gamma$, as in \cite{Bernabo2024} and in Equation \ref{eq:RVs}, which resulted in $\dot{V}_\gamma$=(0.97 $\pm$ 1.49) $\cdot$10$^{-10}$ km s$^{-1}$, compatible with 0 within the errorbars and with the result of Section \ref{sec:complete_model} in Table \ref{tab:model_parameters}. With 17 fitting parameters, the BIC value of this fit is $\simeq$ 1898 and AIC$\simeq$ 1828.

\subsection{Periastron precession model}
\label{sec:periastron}
First of all, we calculated the expected contributions of the apsidal motion rate as (see Equations 1-5 and the Appendix in~\cite{Bernabo2024} for more details), where the orbital periastron angle varies lineary with time: $\omega(t)$=$\omega_0 + \dot{\omega}\cdot(t-T_0)$ and where the rate of apsidal motion $\dot{\omega}$ is the sum of the General Relativity, tidal and rotational contributions: 
\begin{equation}
    \dot{\omega} = \frac{d\omega}{dt} = \frac{d\omega_\mathrm{GR}}{dt} + \frac{d\omega_\mathrm{tidal}}{dt} + \frac{d\omega_\mathrm{rot}}{dt} .
    \label{eq:wdot_tot}
\end{equation}
Such contributions are explicitly written in Appendix A of \cite{Bernabo2024}. Here we calculate them for WASP-43b and they are reported in Table~\ref{tab:wdot_contributions}. Note that the planetary contributions are computed assuming synchronous rotation of the planet ($P_{orb}$/$P_{rot,p}$ = 1) and the two extreme cases for the planetary internal structure: the Love number of the planet $\Lovep$=0.01 and $\Lovep$=1.5. Specifically, $\Lovep$=0.01 represents an extreme case where the planet behaves almost like a point mass with minimal deformation, suggesting a highly rigid body with negligible tidal bulging. On the other hand, $\Lovep$=1.5 stands for a fluid-like, homogeneous body that experiences significant deformation under tidal forces. The main contribution is the planetary one (see Equation 8 and relative discussion in \citealp{Ragozzine2009}) and the total periastron precession rate ranges then from $\simeq$3$^{\prime\prime}$ to $\simeq$146$^{\prime\prime}$, respectively in the case of $\Lovep$=0.01 and $\Lovep$=1.5, while the period of such a motion ranges from $\simeq$1037 to $\simeq$24 years. \\
\begin{table}[]
    \centering
    \caption{Contributions to the periastron precession rate.}
    \begin{tabular}{c c c}
    \hline \hline
    $\Lovep$ & 0.01 & 1.5 \\ 
    \hline 
    $\dot{\omega}_{GR}$ & 2.22 $^{\prime\prime}$ d$^{-1}$ & 2.22 $^{\prime\prime}$ d$^{-1}$ \\
    $\dot{\omega}_{tid, \star}$ & 0.23$^{\prime\prime}$ d$^{-1}$ & 0.23 $^{\prime\prime}$ d$^{-1}$ \\
    $\dot{\omega}_{rot, \star}$ & 0.02 $^{\prime\prime}$ d$^{-1}$ & 0.02 $^{\prime\prime}$ d$^{-1}$\\
    $\dot{\omega}_{tid, p}$ & 0.90$^{\prime\prime}$ d$^{-1}$ & 134.28 $^{\prime\prime}$ d$^{-1}$ \\
    $\dot{\omega}_{rot, p}$ & 0.06$^{\prime\prime}$ d$^{-1}$ & 8.98 $^{\prime\prime}$ d$^{-1}$ \\
    Total $\dot{\omega}$ & 3.42 $^{\prime\prime}$ d$^{-1}$ & 145.72 $^{\prime\prime}$ d$^{-1}$\\
    Period & 1037.4 yrs & 24.35 yrs \\
    \hline
    Measured $\dot{\omega}$ & \multicolumn{2}{c}{(602.98$\pm$29.79) $^{\prime\prime}$ d$^{-1}$} \\
    \hline
    \end{tabular}
    \tablefoot{Expected gravitational, rotational and tidal contribution to the periastron precession in the extreme Love number cases (mass-point and homogeneous planet) and the corresponding apsidal motion period, assuming synchronous rotation for the planet. The value of $\dot{\omega}$ measured in this Section is reported for comparison.}
    \label{tab:wdot_contributions}
\end{table}

\begin{figure}
    \centering
    \includegraphics[width=0.5\textwidth]{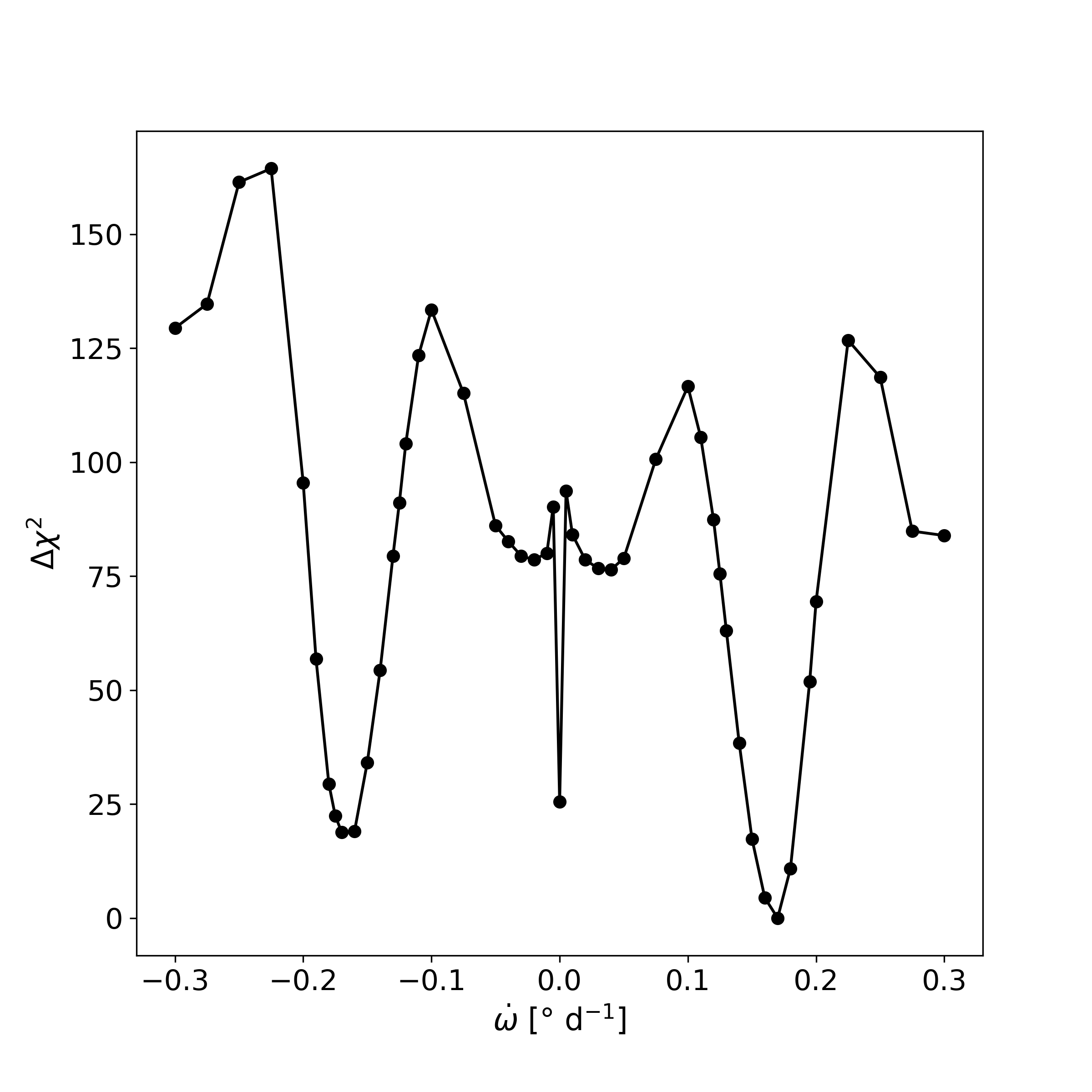}
    \caption{ Results of the semi-grid analysis, with fixed periastron precession rate at 48 different values and the $\chi^2$ of the corresponding fit (relative to the minimum $\chi^2$). }
    \label{fig:semi-grid}
\end{figure}

\noindent As in the case of WASP-19Ab, we fitted a precessing orbit in RVs and TTVs, using Equations 7 and 8 of \cite{Bernabo2024} for the RV fitting and Equations 7 and 8 of \cite{Csizmadia2019} for transits and occultations. As in \cite{Bernabo2024}, we initially applied the semi-grid method, which involves fixing one parameter (the periastron precession rate) at several distinct values while systematically varying other parameters. It's particularly useful for isolating the impact of specific factors in complex models. In the fitting, we kept fixed $\dot{\omega}$ at 48 different values between -0.30 and +0.30$^\circ$ d$^{-1}$, focusing in particular around $\dot{\omega}$=0 $^\circ$ d$^{-1}$. Figure \ref{fig:semi-grid} shows the fit results, with the corresponding chi-squared values. Many local minima appear, as a mirror of the complexity of the problem. \\

\noindent This preliminary analysis allowed us to identify key trends and then proceed to free $\dot{\omega}$ for fitting, leading to a comprehensive analysis. We obtained a periastron precession rate of $\dot{\omega}$=(0.1675$\pm$0.0083)
$^\circ$ d$^{-1}$ = (602.98$\pm$29.79)  
$^{\prime \prime}$ d$^{-1}$. This values corresponds to the global minimum found in Figure \ref{fig:semi-grid}. The BIC value of such a fit is $\simeq$1886 and AIC$\simeq$ 1816. We also want to highlight that  $\dot{\omega} \simeq$603 $^{\prime \prime}$ d$^{-1}$ is well above the highest limit of $\simeq$145$^{\prime \prime}$ d$^{-1}$ calculated for $\Lovep$=1.5. The implications of this result will be discussed in Section \ref{sec:discussion}.

\subsection{Complete model}
\label{sec:complete_model}

\indent Finally, the RV and TTV fitting model of \cite{Bernabo2024} has been implemented with a joint fit of an RV long-term acceleration in the RV trend, and tidal decay and periastron precession. The complete RV equation is:
\begin{equation}
\label{eq:RVs}
\begin{split}    
    V(t)= V_{i,instr} + V_\gamma + \dot{V_\gamma} \cdot(t-t_0) +\delta V(t) +\\
    + K(t) \cdot [e\cdot \cos(\omega(t)) + \cos(v(t)+\omega(t))],
\end{split}
\end{equation}
where $V_{i,instr}$ is the instrumental offset ($i$ indicates the ID number of the dataset, as in Table \ref{Tab:RV_data}), $V_\gamma$ is the systematic velocity of the barycenter of the exoplanetary system, $\dot{V_\gamma}$ is the long-term acceleration, $\delta V$ is apparent contribution due to the distorted stellar shape (see Equation 10 and relative discussion in \citealp{Bernabo2024}). $v(t)$ is the angle describing the true anomaly of the orbit. $\omega(t)$ and $K(t)$ represent the time-changing functions of the periastron angle and of the RV semi-amplitude, respectively. The first one varies linearly with time due to the apsidal motion: $\omega(t)$=$\omega_0 + \dot{\omega}\cdot(t-T_0)$. The second one is not a fitting parameter, but is derived from the linear change in the anomalistic period $P_a(t)$, as described in Section \ref{sec:tidaldecay}.

\indent We implemented the TTV fitting model of \cite{Csizmadia2019} and \cite{Bernabo2024} by allowing both tidal decay and periastron precession to be fitted at the same time. In particular, transit mid-times are now fitted as:
\begin{equation}
\begin{split}
\label{eq:TTVs}
    T_{tr}(t) = T_0 + N_{tr} \cdot P_s(t) + \frac{1}{2} N_{tr}\cdot(N_{tr}-1) \cdot \dot{P}_s \cdot P_s(t) + \\ - \frac{e}{\pi}\cdot P_a(t)\cdot \cos(\omega(t)), 
    \end{split}
\end{equation}
where $N_{tr}$ is the transit (or cycle) number, as in the first column of Table
\ref{Tab:transits1}. $P_s(t)$ represents the time-changing sidereal period, which is derived from the anomalistic period with their relation $P_s(t)$=$P_a(t)/(1-\dot{\omega}/n(t))$, where $n(t)$ is the mean motion and $P_a(t)$=$P_{a,0} + \dot{P}_a(t-T_0)$. 
The first two terms of the equation represent the linear fit, the third one is the quadratic fit (tidal decay) and the last one is the periastron precession. Occultations are fitted in the same way, substituting $N_{tr}$ with $N_{occ}$, with a positive sign before the last term and with an additional +$P_s(t)/2$ term to account for the half orbit that the planet has travelled between the transit and the occultation event.\\


\begin{table*}[]
    \centering
    \caption{Results of the complete fit.}
    \tiny
    \resizebox{\textwidth}{!}{
    \begin{tabular}{l l l c l}
    \hline 
    \hline 
        Parameter & Symbol [Units] & Fit & Value & Priors\\
        \hline
        {\bf Assumed parameters} & & & &\\
        Stellar rotation & {\it V $\sin$ i}  [km s$^{-1}$] & & 2.26 \\
        Stellar mass & {\it M}$_\mathrm{\star}$ [M$_{\sun}$] & & 0.717\\
        Planet-to-star radius ratio & {\it r}$_\mathrm{p}$/{\it R}$_\mathrm{\star}$ & & 0.15940 \\
        Semi-major axis & {\it a}$_\mathrm{0}$ [AU] & & 0.01526 \\
        First transit epoch & {\it T}$_0$ [$\BJDTDB$-2~454~000]
        & & 5528.86823 \\
        \\
        {\bf Adjusted parameters}\\
        Jitters \hspace{3.2cm} ID 1 & {\it jit}$_1$ [m s$^{-1}$] & RV & 2.1 \\
        \hspace{3.93cm} ID 2 & {\it jit}$_{2}$ [m s$^{-1}$] & RV & 0.0 \\
         \hspace{3.93cm} ID 3 & {\it jit}$_{3}$ [m s$^{-1}$] & RV & 11.4 \\ 
         \hspace{3.93cm} ID 4 & {\it jit}$_{4}$ [m s$^{-1}$] & RV & 13.8  \\
         \hspace{3.93cm} ID 5 & {\it jit}$_{5}$ [m s$^{-1}$] & RV & 18.4 \\
         \hspace{3.93cm} ID 5 & {\it jit}$_{6}$ [m s$^{-1}$] & RV & 17.1  \\
        \hspace{3.93cm} ID 6 & {\it jit}$_{7}$ [m s$^{-1}$] & RV & 4.5 \\
        \hspace{3.93cm} ID 7 & {\it jit}$_{8}$ [m s$^{-1}$] & RV & 6.1 \\
         \hspace{3.93cm} Transits & {\it jit}$_\mathrm{tr}$ [d] & Tr & 0.00104 \\
         \hspace{3.93cm} Occultations & {\it jit}$_\mathrm{occ}$ [d] & Occ & 0.00023 \\
         
        {\bf Fitted parameters}\\
        Offset velocity of the system & {\it V}$_\gamma$ [km s$^{-1}$] & RV & -3.602 $\pm$ 0.013 & $\mathcal{U}$ [-$\infty$,+$\infty$]\\
        Long-trend acceleration & $\dot{V}_\gamma$ [km s$^{-2}$] & RV & (1.05 $\pm$ 1.48) $\cdot$ 10$^{-10}$ & $\mathcal{U}$ [-$\infty$,+$\infty$]\\
        RV semi-amplitude & $K_0$ [km s$^{-1}$] & RV & 0.5543 $\pm$ 0.0017 & $\mathcal{U}$ [-$\infty$,+$\infty$]\\
        Offset instrumental \hspace{1.7cm} ID 2 
        & {\it V}$\mathrm{_{instr,2}}$ [km s$^{-1}$] & RV & -0.00011 $\pm$ 0.01731 & $\mathcal{U}$ [-$\infty$,+$\infty$]\\
        velocity \hspace{2.95cm} ID 3 & {\it V}$\mathrm{_{instr,3}}$ [km s$^{-1}$] & RV & +0.017 $\pm$ 0.014 & $\mathcal{U}$ [-$\infty$,+$\infty$]\\
        \hspace{3.93cm} ID 4 & {\it V}$\mathrm{_{instr,4}}$ [km s$^{-1}$] & RV & +0.0011 $\pm$ 0.0192 & $\mathcal{U}$ [-$\infty$,+$\infty$]\\
        \hspace{3.93cm} ID 5 & {\it V}$\mathrm{_{instr,5}}$ [km s$^{-1}$] & RV & +0.038 $\pm$ 0.029 & $\mathcal{U}$ [-$\infty$,+$\infty$]\\
        \hspace{3.93cm} ID 6 & {\it V}$\mathrm{_{instr,5}}$ [km s$^{-1}$] & RV & +0.0035 $\pm$ 0.029 & $\mathcal{U}$ [-$\infty$,+$\infty$]\\
        \hspace{3.93cm} ID 7 & {\it V}$\mathrm{_{instr,6}}$ [km s$^{-1}$] & RV & +0.00082 $\pm$ 0.02888 & $\mathcal{U}$ [-$\infty$,+$\infty$]\\
        \hspace{3.93cm} ID 8 & {\it V}$\mathrm{_{instr,7}}$ [km s$^{-1}$] & RV & +0.0023 $\pm$ 0.0425 & $\mathcal{U}$ [-$\infty$,+$\infty$]\\
        & $\esin$ & RV, Tr/Occ & -0.031 $^{+0.010}_{-0.014}$ & $\mathcal{U}$ [-1,+1] \\
        & $\ecos$ & RV, Tr/Occ & -0.024$^{+0.014}_{-0.011}$ & $\mathcal{U}$ [-1,+1]\\
        Anomalistic period & {\it P}$_\mathrm{a,0}$ [d] & RV, Tr/Occ & 0.813792 $^{+0.000015}_{-0.000017}$ & $\mathcal{U}$ (0.81;0.1)\\ 
        Tidal decay rate & $\dot{P}_\mathrm{a}$ [ms yr$^{-1}$] & RV, Tr/Occ & -1.99 $\pm$ 0.50, $\pm$ 0.99 & $\mathcal{U}$ [-$\infty$,+$\infty$]\\
        Periastron precession rate & $\dot{\omega}$ [$^\circ$ d$^{-1}$] & RV, Tr/Occ & 0.1727$^{+0.0083}_{-0.0089}$ \\
        & \hspace{0.25cm} [$^{\prime\prime}$ d$^{-1}$] & & 621.72 $^{+29.88}_{-32.04}$ & \\
        Stellar Love number & {\it k}$_\mathrm{2,\star}$ & RV & 0.020 $\pm$ 0.011 & $\mathcal{N}$ (0.020; 0.004)\\
        Stellar temperature & {\it T}$_\mathrm{eff,\star}$ [K] & RV & 4523.79 $\pm$ 270.51 & $\mathcal{N}$ (4500 K; 100 K)\\
        Planet-to-star mass ratio & {\it q} & RV & 0.00273 $\pm$ 0.00027 & $\mathcal{N}$ (0.00273; 9.9$\cdot$10$^{-5}$) \\ 
        Rotational stellar period$^\dagger$ & {\it P}$_\mathrm{rot,\star}$ [d] & RV & 13.24 $^{+6.11}_{-5.85}$ & $\mathcal{N}$ (15.0; 10.0) \\
        Orbital inclination & {\it i} [$^\circ$] & RV, Tr/Occ & 82.32 $\pm$ 1.08 & $\mathcal{N}$ (82.3$^\circ$; 1.0$^\circ$) \\
        & & & \\
    {\bf Derived parameters} \\
    Eccentricity & {\it e} & & 0.00188 $\pm$ 0.00035, $\pm$ 0.00070, $\pm$ 0.00160 & \\
    Periastron angle & $\omega_0$ [$^{\circ}$] & & 47.09 $\pm$ 23.89 & \\ 
    Sidereal period & {\it P}$_\mathrm{s}$ [d] & & 0.81347489 $\pm$ 9.9e-8 & \\
    RV semi-amplitude increase & $\dot{K}$ [m s$^{-1}$ yr$^{-1}$] & & (1.58 $\pm$ 0.39)$\cdot$ 10$^{-5}$ & \\ 
    \hline 
    \end{tabular}
    }
    \tablefoot{The Table shows the assumed, adjusted, fitted parameters with their priors, and derived parameters 
    The third column states whether the fitted parameter is used to model radial velocity ("RV") and/or mid-transit and mid-occultation times ("Tr/Occ"). Errorbars are reported with 1$\sigma$ confidence level for all parameters, while for $e$, $\dot{\omega}$ and $\dot{P}_a$ they are reported also with 2$\sigma$ confidence level. For $e$ we also report the 5$\sigma$ confidence level to strengthen the findings. \\
    $^\dagger$ The stellar rotational period is fitted as in Equation (10) in \cite{Bernabo2024}.
    }
    \label{tab:model_parameters}
\end{table*}

\begin{figure}
    \centering
    \includegraphics[width=0.5\textwidth]{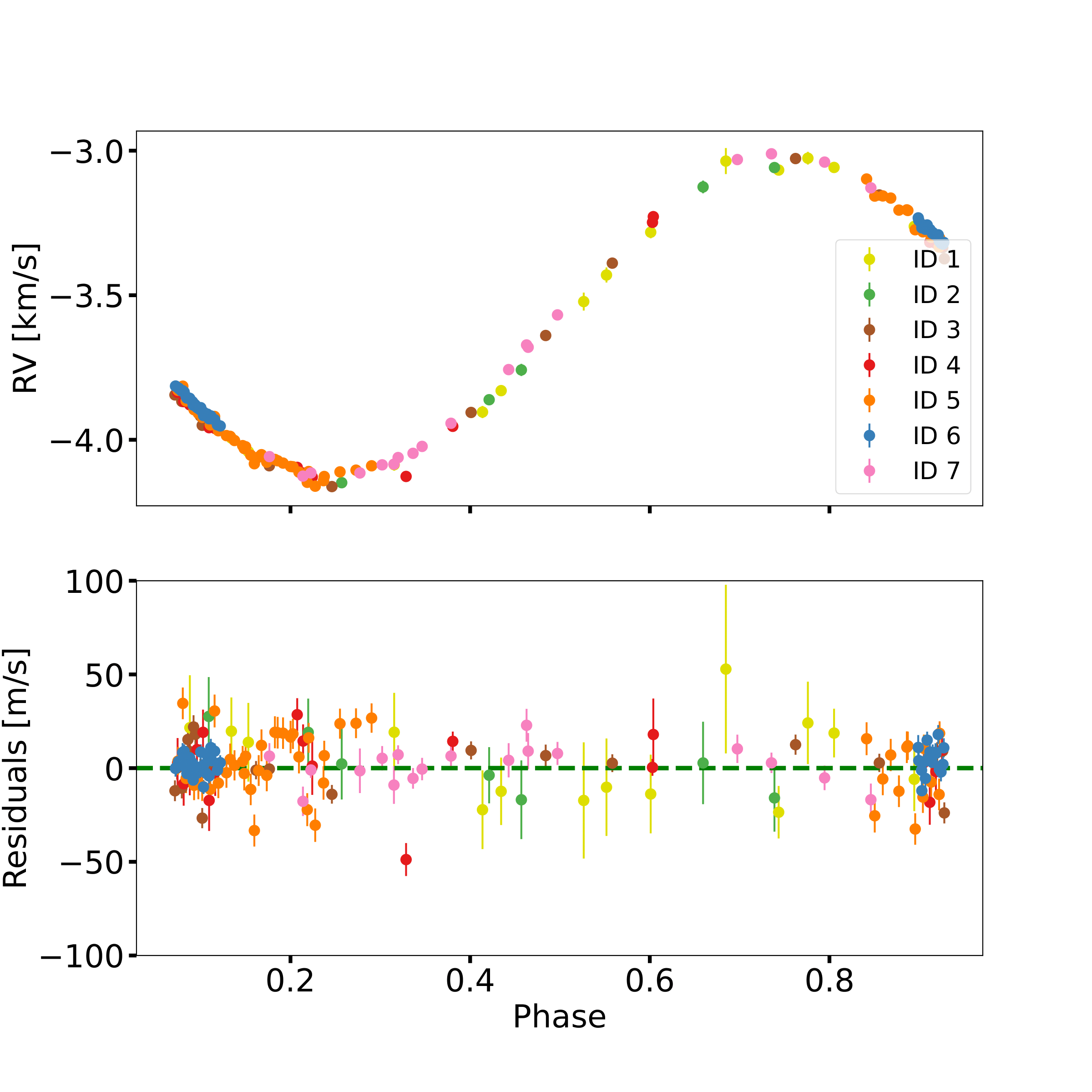}

    \caption{
    {\it Top plot}: Phase-folded RV data with periastron precession rate of 0.1727 $^{\circ}$d$^{-1}$ and residuals, with a standard deviation of $\simeq$ 13.5 m s$^{-1}$.}
    \label{fig:residuals_RVs}
\end{figure}

\begin{figure}
    \centering
    \includegraphics[width=0.5\textwidth]{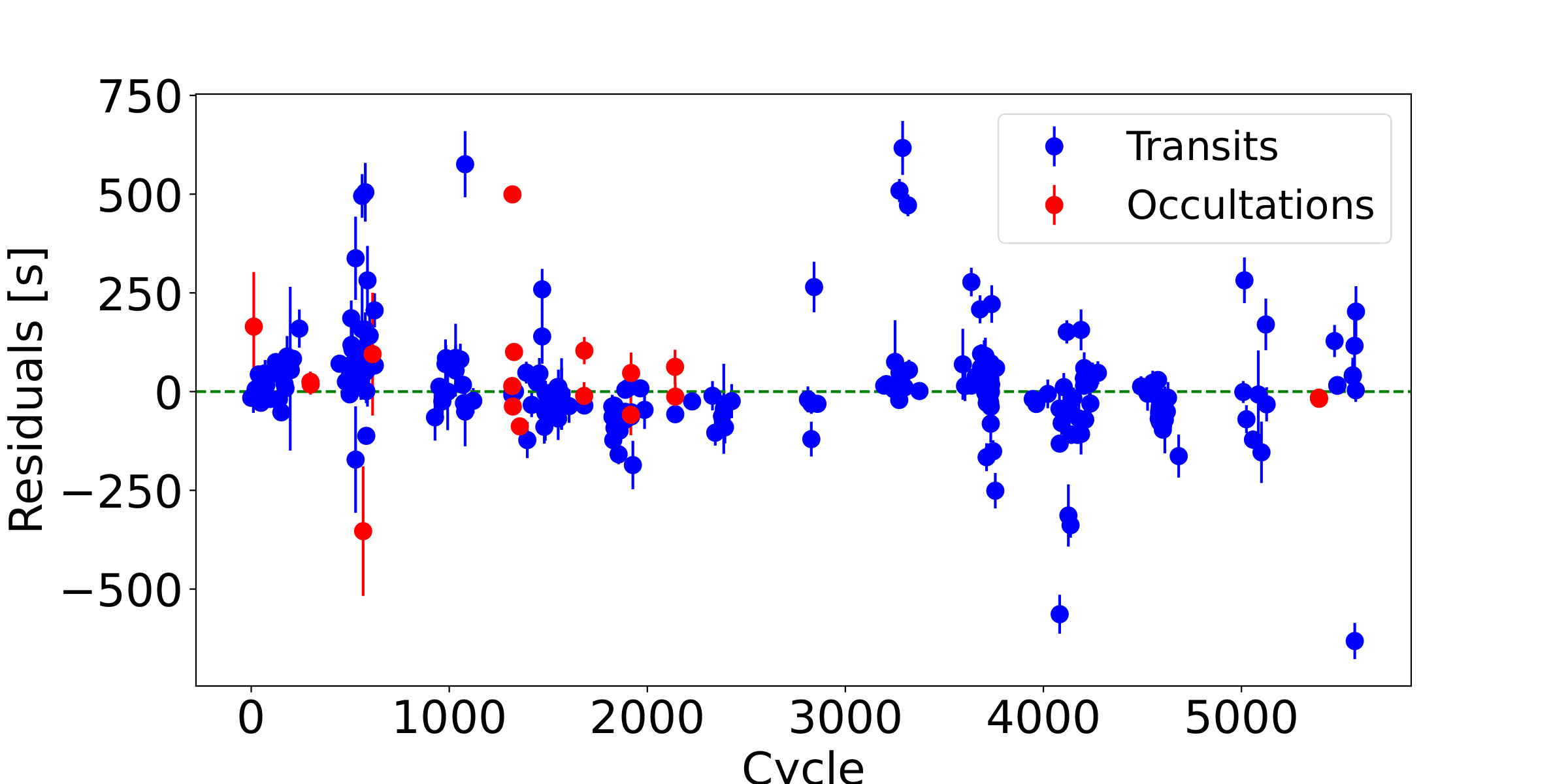}
    \caption{
    Residuals of transit and occultation mid-times, with a standard deviation of $\simeq$ 127.4 and 155.6 s, respectively.}
    \label{fig:residuals_TTVs}
\end{figure}
\begin{figure}
    \centering
    \includegraphics[width=0.5\textwidth]{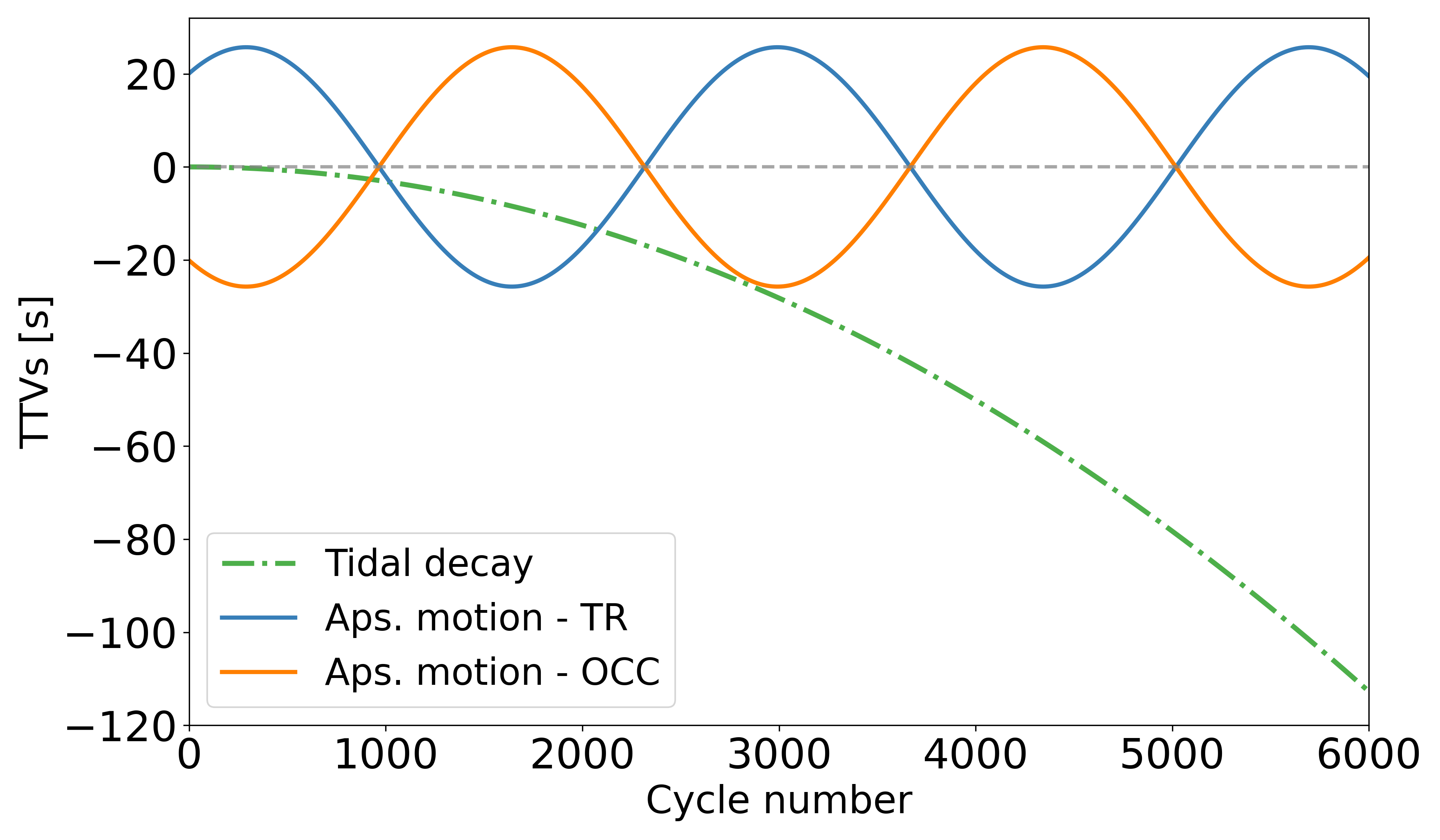}
    \caption{Transit timing variations caused by the measured tidal decay (dashed-dotted line in green) and apsidal motion contributions (respectively in blue for transit mid-times and orange for occultation mid-times) for the $\simeq$10 years of transit and occultation data used in this paper.}
    \label{fig:Tiddecay_apsmotion}
\end{figure}

\noindent The results of the joint fit (RVs, transit and occultation mid-times), including apsidal motion, tidal decay and long-term acceleration, can be found in Table \ref{tab:model_parameters}, reporting the median values and 1$\sigma$ uncertainty for each fitting parameter. For  $\dot{\omega}$ and $\dot{P}_a$ we also report the 2$\sigma$ confidence levels. As discussed in Section \ref{sec:eccentricity}, since the eccentricity is a key parameter for our study and an eccentric orbit is a precondition for periastron precession to take place, we report 2 and 5$\sigma$ errorbars for $e$. Figures \ref{fig:residuals_RVs} shows the phase-folded RV data and residuals of this complete model. Figure \ref{fig:residuals_TTVs} shows the mid-transit and mid-occultation timing residuals, which show no visible deviation from a zero-trend.\\
\indent We detect a periastron precession rate of $\dot{\omega}$=(0.1727$^{+0.0083}_{-0.0089}$)$^\circ$ d$^{-1}$ = (621.72 $^{+29.88}_{-32.04}$)$^{\prime\prime}$ d$^{-1}$, compatible with the result in Section \ref{sec:periastron}. As in \cite{Bernabo2024}, a small number of the MCMC chains converged to the same value but with negative sign, showing how complex the convergence of such a fitting is. Within the errorbars, the positive $\dot{\omega}$ rate agrees with the result of the periastron precession fit in Section \ref{sec:periastron}. The tidal decay rate is $\dot{P_a}$=(-1.99 $\pm$ 0.50) ms yr$^{-1}$= (-0.00195 $\pm$ 0.00050) s yr$^{-1}$, which is in agreement with the result of the tidal decay fit of Section \ref{sec:tidaldecay}. We attribute such small errorbars ($\simeq$5\% for $\dot{\omega}$ and $\simeq$ 27\% for $\dot{P_a}$) to the fact that for the first time we fit simultaneously RVs and TTVs and to the long and wide baseline with respect to all previous papers that investigated tidal decay. \\  
\indent Figure \ref{fig:Tiddecay_apsmotion} shows such contributions in TTVs over time. The contributions of apsidal motion in transit and occultation mid-times has opposite phase, as in Equation \ref{eq:TTVs}. In certain regions, tidal decay and apsidal motion can cancel each other, making it hard to distinguish the two if a long enough baseline of observations is not available. Even if the amplitude of the contribution of apsidal motion is smaller than the errorbars of some of the transit and occultation mid-times, the detection of apsidal motion is possible due to the combination of RV and TTV analysis and the length of the database ($\simeq$ 10 years) and the number of datapoints (471 transits, occultations and out-of-transit RVs) we make use of.\\
\indent This rate of tidal decay corresponds to an increase of the RV semi-amplitude $\dot{K}$ = (1.58$\pm$ 0.39) $\cdot$ 10 $^{-5}$ m s$^{-1}$ yr$^{-1}$. We do not detect a linear trend (long-term acceleration $\dot{V}_\gamma$), supporting the fact that there is no unseen third body in the system (see Section \ref{sec:discussion} for a complete analysis). The apsidal motion has a period of $U$=$2 \pi / \dot{\omega} \sim $5.7 years, meaning that the periastron angle $\omega$ has completed two cycles within the length of our observations. As shown in Table \ref{tab:BIC}, with 19 fitting parameters, the BIC value is $\simeq$1849 and AIC is $\simeq$1771. Statistically this is the favoured model, even if it has the highest numbers of parameters. With the inclusion of the jitters - which are adjusted post-fitting - the joint fit has a reduced $\chi^2$ of $\simeq$ 0.97. \\

\begin{figure}
    \centering
    \includegraphics[width=0.37\textwidth]{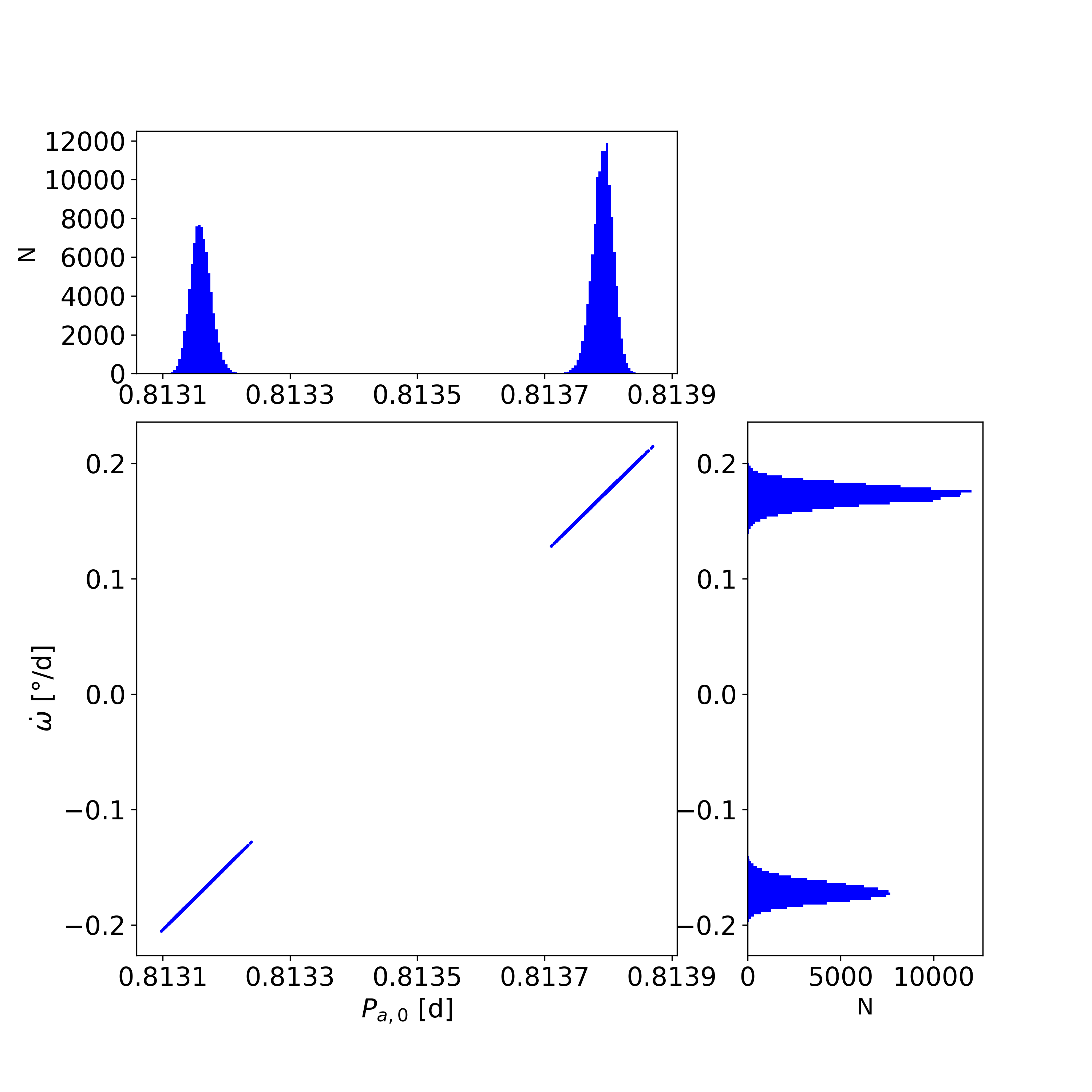}
    \includegraphics[width=0.37\textwidth]{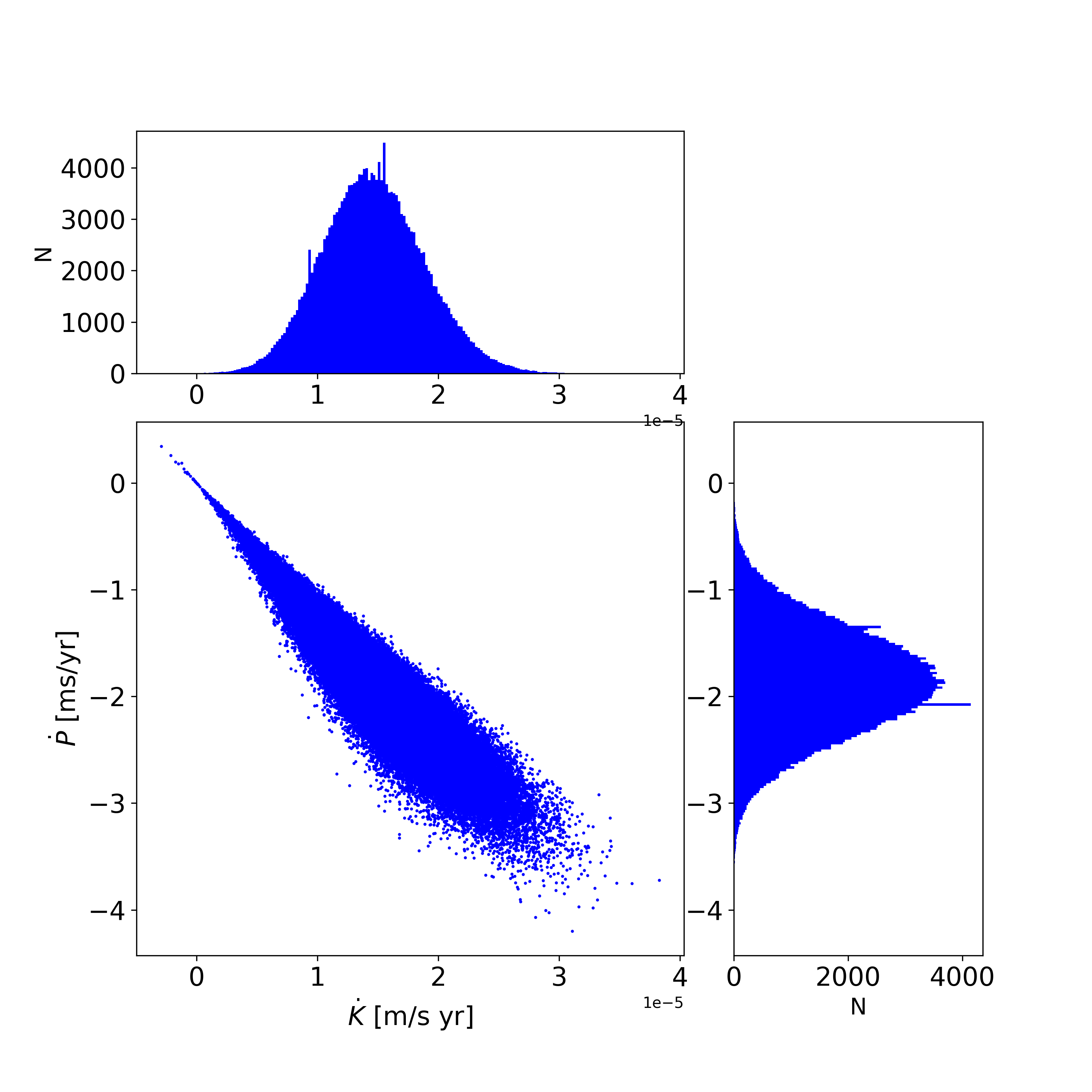}
    \includegraphics[width=0.37\textwidth]{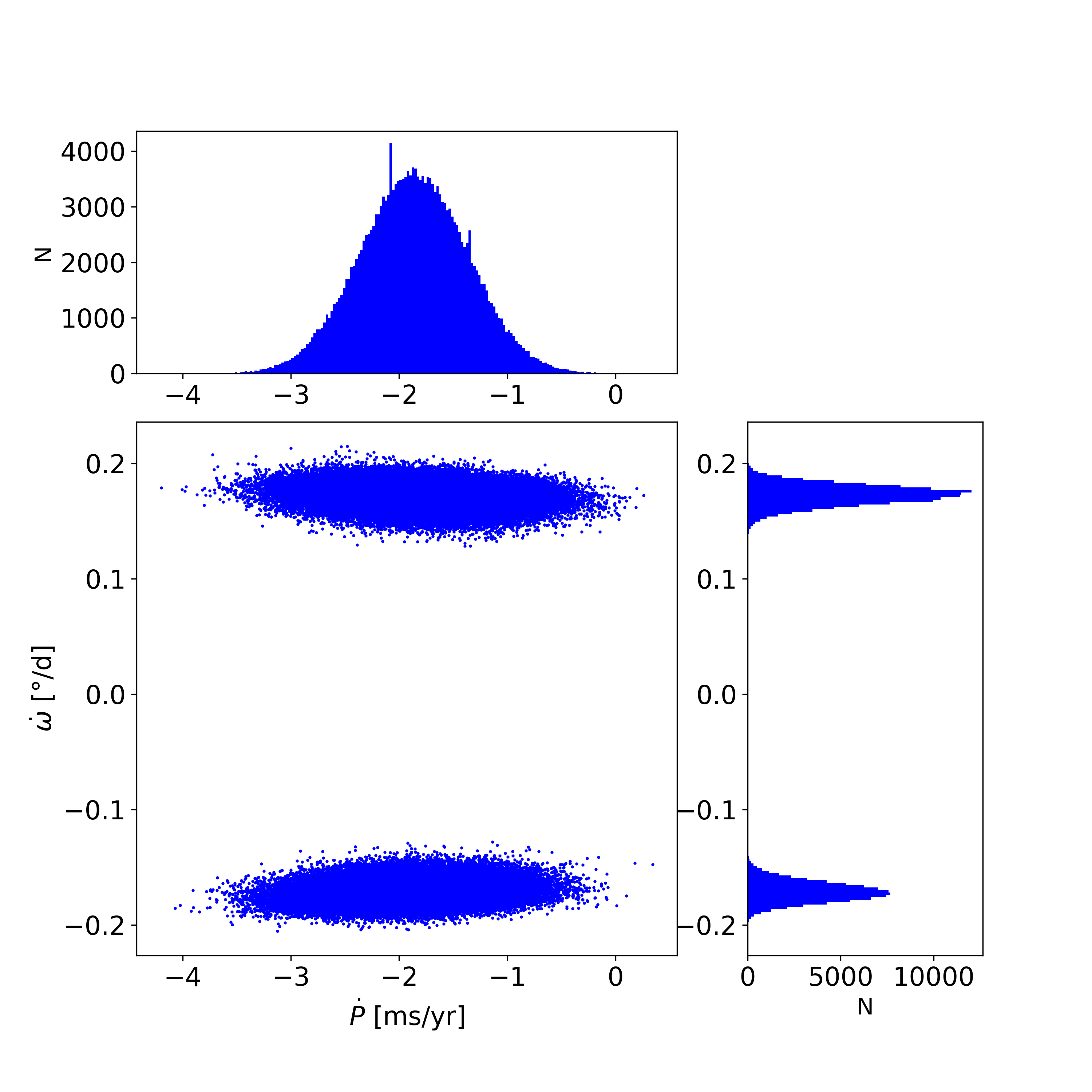}
    \caption{Posterior distribution of the parameters, for all ten chains, excluding the DE-MCMC steps in the burn-in phase.\\
    {\it Top plot} The correlation between the anomalistic period and the periastron precession rate. A small number of chains were stuck in a local minimum with the same value as the detected $\dot{\omega}$ value, but negative sign. Due to the strong correlation between $\dot{\omega}$ and $P_{a,0}$, also the latter shows a double peaked posterior distribution. \\
    {\it Middle plot} The correlation between the tidal decay rate ($\dot{P_a}$, a fitting parameter) and the change in the RV semi-amplitude $\dot{K}$, which is derived from its relation with the tidal decay rate $\dot{P_a}$. The spikes in the posterior distributions denote particularly investigated regions in the parameter space and the fact that they chains may get stuck in local minima before reaching the global minimum.\\
    {\it Low plot} The correlation between periastron precession rate and tidal decay rate, showing that the two parameters are independent. The two opposite solutions of $\dot{\omega}$ are visible}
    \label{fig:posterior_distr}
\end{figure}

\begin{table}[]
    \centering
    \caption{Model comparison.}
    \begin{tabular}{l c c c c}
    \hline \hline
        Model & k & BIC & AIC & Section \\
        \hline
        One planet, circular & 14 & 2190 & 2132 & \ref{sec:simple_models}\\
        One planet, eccentric & 16 & 1855 & 1789 & \ref{sec:simple_models}\\
        Tidal decay & 17 & 1946 & 1876 & \ref{sec:tidaldecay}\\
        Long-term acc. & 17 & 1898 & 1828 & \ref{sec:longterm}\\
        Apsidal motion & 17 & 1886 & 1816 & \ref{sec:periastron} \\
        Complete & 19 & 1849 & 1771 & \ref{sec:complete_model}\\
        \hline
    \end{tabular}
    \tablefoot{Comparison between the fitting models considered in this study: number of parameters $k$, BIC and AIC values for each of them.}
    \label{tab:BIC}
\end{table}
\noindent Figure \ref{fig:posterior_distr} shows the posterior distributions of the most critical fitting parameters (considering only the last 2$\cdot$10$^{4}$ steps of the DE-MCMC procedure, excluding the burn-in phase). In the top plot, we show the relation between the anomalistic period and the periastron precession rate. These two make one of the more correlated and degenerate couple of parameters in our model due to the well-known equation: $P_s (t) = P_a(t)\cdot(1-\dot{\omega}/n(t))$, where $P_s (t)$, $P_a (t)$ and $n(t)$ are time-varying parameters. The middle plot shows the correlation between the tidal decay rate and the rate of the RV semi-amplitude change $\dot{K}$, as in Equation \ref{eq:Kdot}. The lower plot shows the posterior distributions of $\dot{\omega}$ and $\dot{P}_a$, which appear not to be correlated.\\

\noindent The high periastron precession rate of $\dot{\omega}$=(0.1727$^{+0.0083}_{-0.0089}$)$^\circ$ d$^{-1}$ has never been observed in any exoplanetary system. Even with similar orbital parameters to WASP-18Ab (\citealp{Csizmadia2019}) and WASP-19Ab (\citealp{Bernabo2024}), it is $\simeq$20 times bigger with respect to the first one and almost three times bigger than the latter. In the next Section, we will discuss some possible causes of such a high detected rate of apsidal motion.

\section{Discussion}
\label{sec:discussion}

\subsection{WASP-43b is a fast rotator}

\begin{figure}
    \centering
    \includegraphics[width=0.5\textwidth]{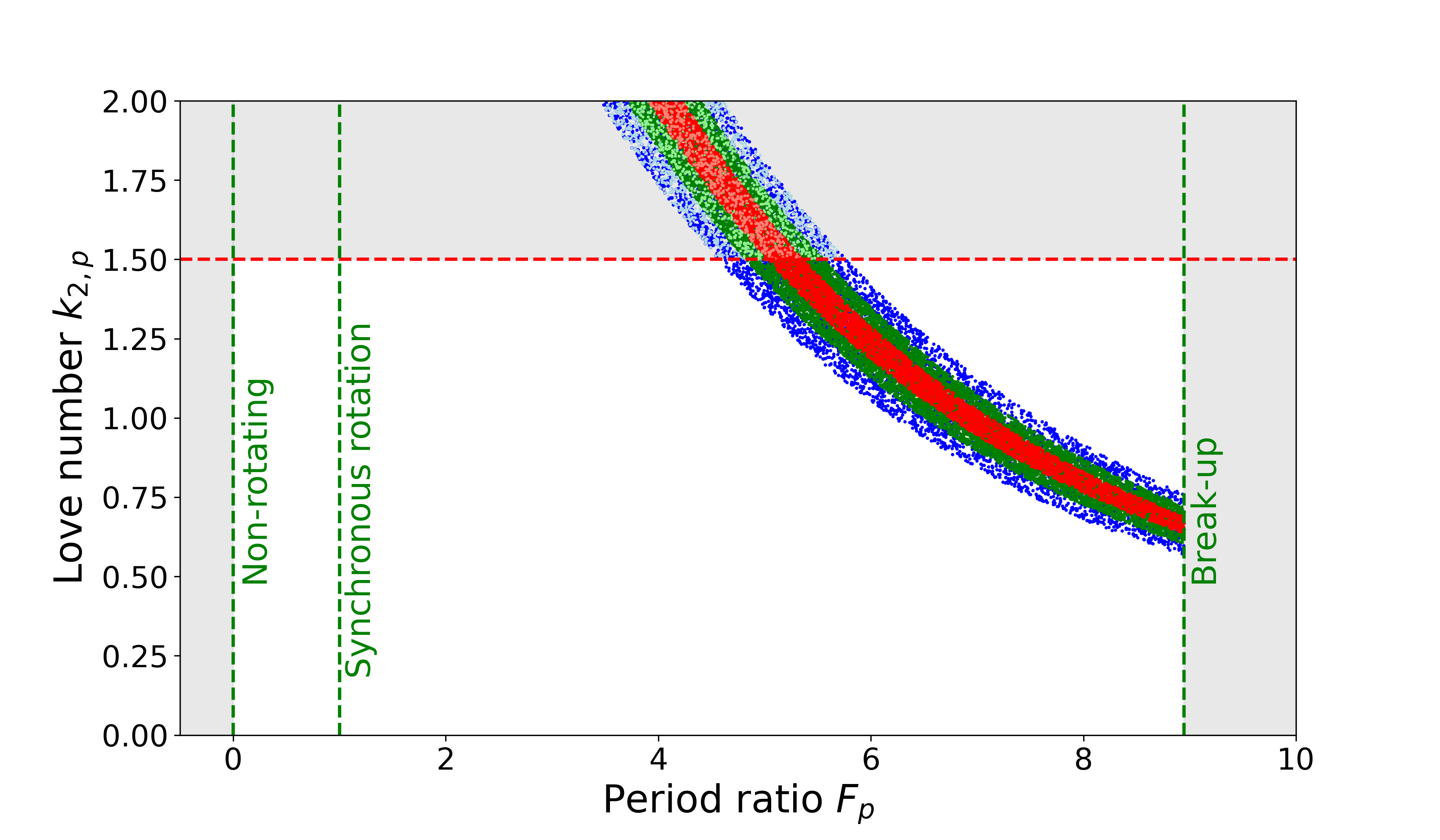}
    \caption{Couples of period ratio and Love number of the planet that solve Equations \ref{eq:omega_dot_tidal} and \ref{eq:omega_dot_rot}. The blue and light blue points correspond to a combination of ($F_p$, $\Lovep$) that satisfies the observed periastron precession rate within its 1, 2 and 3$\sigma$ errorbars (red, green and blue areas, respectively). 
    The red horizontal line represents the physical limit where the body is homogeneous. From left to right, the three vertical lines highlight a non-rotation planet ($\Porb$/$P_\mathrm{rot,p}\rightarrow$0), synchronous rotation ($P_\mathrm{rot,p}$=$\Porb$) and the break-up limit ($P_\mathrm{rot,p}\simeq$8.95$\Porb$, corresponding to $P_\mathrm{rot,p}\simeq$2.18 hours). The grey shaded regions are unphysical.}
    \label{fig:Fp_k2p}
\end{figure}

Once the periastron precession rate is measured, we subtracted the General Relativity term and the tidal and rotational components relative to the star from the observed value (see Equation \ref{eq:wdot_tot}). As described in Appendix A of \cite{Bernabo2024}, the remaining terms are the tidal and rotational contributions of the planet, whose first orders approximation are:
\begin{equation}
\begin{split}
    \frac{d\omega_\mathrm{tidal,p}}{dt} \simeq \frac{15}{{2}} \frac{n}{(1-e^2)^5} \cdot \frac{\Ms}{\mpl} \Lovep \left(\frac{R_\mathrm{p}}{a}\right)^5  \left( 1 + \frac{3}{2}e^2+ \frac{1}{8} e^4 \right),
    \label{eq:omega_dot_tidal}
\end{split}
\end{equation}
and
\begin{align}
\begin{split}
    \frac{d\omega_\mathrm{rot,p}}{dt} \simeq 
    \frac{1}{{2}}
    \frac{n}{(1-e^2)^2}  \left(\frac{P_a}{P_\mathrm{rot,p}} \right)^2 \Lovep \left( \frac{R_\mathrm{p}}{a} \right)^5 \left(1+ \frac{\Ms}{\mpl} \right),
    \label{eq:omega_dot_rot}
    \end{split}
\end{align}
where $\Ms$ and $\mpl$ the stellar and planetary masses, $\rp$ the planetary radius (see also \citealp{Sterne1939}), $\Loves$ and $\Lovep$ the fluid Love numbers of star and planet and $\Prots$ and $\Protp$ the stellar and planetary rotation period around their own axis. The other parameters have been defined already. \\
\indent These equations depend on two unknowns: $\Lovep$ and $\Protp$. To solve the equations for $\Lovep$, we have to assume $\Protp$. The two extreme cases are when the planet rotates at the breakup velocity - corresponding to $\Protp \simeq$ 2.18 hours - and when it is not rotating ($F_p$=$P_a$/$P_{rot,p} \rightarrow$ 0).\\
\indent The observed apsidal motion rate is four times the maximum rate in Table~\ref{tab:wdot_contributions}, which is calculated assuming synchronization of the planetary rotational and orbital period $\Protp$=$\Porb$ (or $F_p$=1). With this assumption, the Love number we derive is way higher than the physical limit of 1.5, which corresponds to a homogeneous body. No physical solution to the previous equations is found with such an assumption. Therefore, we relaxed the assumption and calculated solutions for the parameter couple $\Lovep$ and $\Protp$, which are plotted in Figure~\ref{fig:Fp_k2p}. The red, green and blue areas represent all such solutions, within 1, 2 and 3$\sigma$ errorbars for $\dot{\omega}$, respectively. The grey area above $\Lovep$=1.5 contains unphysical solutions (representing interior structure models with an envelope denser than the core). The three vertical lines denote respectively the cases of non-rotating planet, synchronous rotation and the extreme scenario of break-up velocity. \\
\indent We can therefore provide a range for $\Lovep$ depending on $\Protp$. Figure~\ref{fig:Fp_k2p} suggests that WASP-43b is a fast rotator with a lower limit on $\Protp\gtrsim$ 4.26 hours (for $F_p\simeq$4.58, the lowest 3$\sigma$ limit visible in the Figure) to and on $\Lovep \gtrsim$ 2.18 hours (the lowest 3$\sigma$ limit in correspondence of $F_p \simeq $ 8.95, when the planet is rotating at the break up velocity).\\

\noindent The fact that WASP-43b may rotate faster than expected could be caused by the impact of a massive object. A significant collision with a large celestial body could have imparted additional angular momentum to the planet, thereby increasing its rotational speed, as it is believed that Earth's relatively rapid rotation and the formation of the Moon resulted from a colossal impact with a Mars-sized object in the early stages of its development (see \citealp{Daly1946} and reference therein). Such impacts are not uncommon in the chaotic environments of young planetary systems, where collisions can significantly influence the rotational dynamics of planets. \\
\indent Alternatively, processes within the planet's mantle or core, such as vigorous convection or dynamic core movements, can redistribute mass internally, altering the planet's moment of inertia, and potentially leading to changes in its rotation rate over time. \\
\indent In most cases of such a tight orbit, we can reasonably assume that during the lifetime of the system, the orbit has been synchronized due to gravitational tidal forces. As \cite{Rauscher2023} shows, assuming Jupiter's quality factor and its initial rotational state, planets with an orbital period below $\simeq$ 30 days have a synchronization timescale of $\lesssim$ 4 Gyrs and are supposed to be aligned and synchronized. However, the age of WASP-43 is not known. \cite{Scandariato2022} reports an age of 7.4$^{+6.4}_{-5.3}$ Gyr from isochrones, while \cite{Gallet2019}'s  tidal-chronology technique derives a range between 2.4 and 9.2 Gyr and \cite{Hellier2011} calculates the gyrochronological age as 400$^{+200}_{-100}$ Myr from the stellar rotation period. Since the planetary orbit is tidally decaying, this estimate may not be entirely reliable, as the host star could be spun up, making the derived age appear younger than the actual age (see e.g. \cite{Maxted2015}). 
Moreover, the planetary orbit is eccentric with $e$=0.00188$\pm$0.00035, even though the planet orbits very close to the host star. This is the only Ultra-Short Period planet whose orbit is eccentric (see Section \ref{sec:eccentricity}) and which does not appear to have a companion star or planet which can drive the eccentricity as in the case of WASP-19Ab (see Appendix B in \citealp{Bernabo2024}). Then the orbit may have not been circularized by tidal forces yet. This supports the hypothesis of a young system, since the typical circularization timescale has an order of magnitude of 10$^{-3}$ Gyrs (derived from Equation 5 in \citealp{Charalambous2023} assuming Jupiter's quality factor and Love number $k_2$, in agreement with \citealp{Terquem2021}). If the orbit is not circularized yet, it may also not be synchronized yet. \\
\indent The JWST light curve published and analysed by \cite{Bell2024} does not support this hypothesis. If the planet rotates $\simeq$4 times faster than its orbital velocity, there should not be a big temperature gradient between the day side and the night side of the planet. However, \citealp{Bell2024} detected a day-to-night temperature contrast of $\simeq$700 K. 
On the other hand, \cite{Lesjak2023} recovered a line broadening in the atmosphere, corresponding to an equatorial velocity of 11.7$^{+2.8}_{-2.1}$ km s$^{-1}$, or equivalently a rotational period of 10.80$^{+2.6}_{-1.9}$ hours. Assuming tidal locking, such a detected velocity is larger than the synchronous rotational velocity. The paper claims that indicative of a super-rotating equatorial jet with a velocity of 5.4$^{+2.8}_{-2.1}$ km s$^{-1}$. Unless the planet rotational axis is tilted with respect to the orbital plane, these recent findings may also be favourable to a faster rotational period than the synchronous one. Moreover, \cite{Stevenson2017} and \cite{Scandariato2022} report offsets of a hotspot measured as ($21.1 \pm 1.8$)$^\circ$ from Spitzer phase curve and ($50^{+30}_{-20}$)$^\circ$ from CHEOPS phase curve, respectively. These results suggest that the rotation rate of the hotspot is not consistent with a tidally locked state (\cite{PennVallis2018}). However, it is unclear whether the rotation rate of the hotspot directly represents the rotation rate of the planet. For a detailed discussion on the relationship between wind rotation rates and the planetary rotation at different pressure levels see \cite{Carone2020}. If the hotspot's rotation rate indeed corresponds to the planetary rotation rate, this would would be an indication that the planet is not tidally locked. More observations are needed in this direction, such as future Rossiter-McLaughlin measurements on the planet to detect the inclination of its rotational axis.\\



\subsection{Presence of a third, unseen body}
\label{sec:perturber}
In this Section we discuss if the observed periastron precession rate can be explained by the presence of a third not yet discovered body in the system, e.g. a companion star or planet.\\
\indent We searched Gaia DR3 (\citealp{Gaia2016,Gaia2022}) for possible co-moving companions to WASP-43, using the same method that revealed probable bound companions to WASP-18A (\citealp{Csizmadia2019}) and WASP-19A (\citealp{Bernabo2024}). Searching to a projected separation of $\simeq$10$^5$ AU (19.2$^{\prime \prime}$), we found no candidates with parallax and proper motion values consistent (even to 10$\sigma$) with those of WASP-43. As far as the Gaia DR3 evidence goes, WASP-43 is an isolated single star.\\

\begin{figure}
    \centering
   \includegraphics[width=0.5\textwidth]{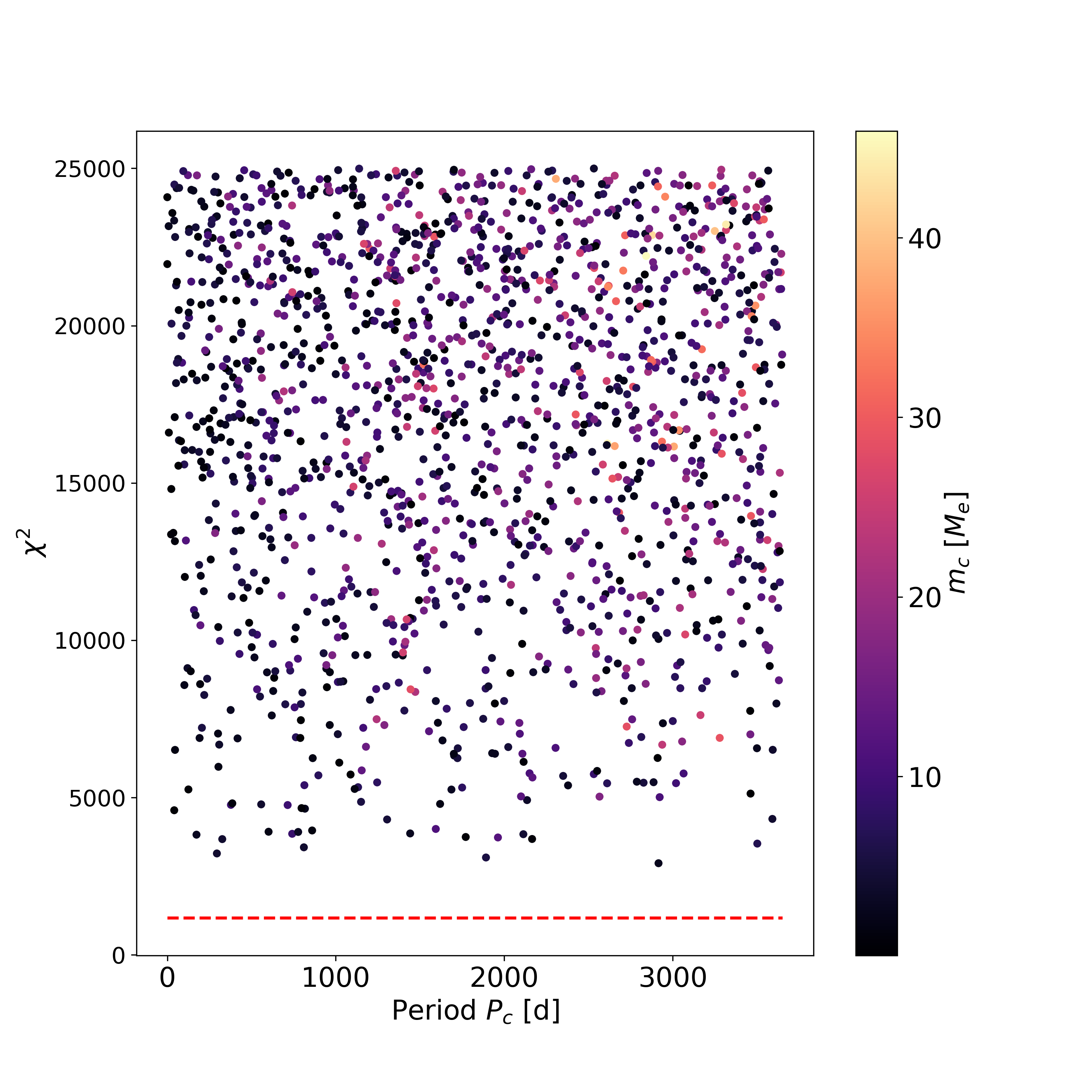}
    \caption{Random candidate planet c fit $\chi^2$ against its orbital period. The dashed horizontal lines correspond to the $\chi^2$ needed for weak evidence of the presence of a second body, according to the BIC criterion (see text for more details). The candidates are coloured by planetary mass.}
    \label{fig:random_2nd_planet}
\end{figure}

\noindent As well as in \cite{Bernabo2024}, we fitted the RV datasets separately with an additional Keplerian signal, beyond that of planet {\it b}, to mimic a hypothetical candidate companion in the system. We "injected" such a candidate in the RV fitting model with different random orbital parameters (orbital period, RV semi-amplitude, $\esin$ and $\ecos$). 
We randomly varied the other orbital parameters as follows: $\esin$ and $\ecos$ between -1.0 and 1.0, the RV semi-amplitude $K_\mathrm{c}$ between 0.1 (the minimum RV errorbar) and 125 m s$^{-1}$ (the maximum RV residual after the subtraction of planet {\it b}). \\
\indent We applied such an analysis to the datasets with a large number of data and well-spread observations in time. We selected datasets with a length of at least a few days: IDs 1, 2, 4 and 8. We also applied it to the whole RV dataset after removing the offsets from each dataset. We investigated 2 $\cdot$10$^8$ random candidates for each separate dataset and 10$^9$ for the complete data collection. The orbital period of such a candidate was constrained not to have collisions with the star (see Equation 6 of \citealp{Bernabo2024}) and below the length of the observations of the corresponding datasets (approximately 187 days for ID 1, 38 days for ID 2, 782 days for ID 4, 18 days for ID 8 and 10 years for the complete RV set). The results are shown in Figure~\ref{fig:random_2nd_planet}: we plot the orbital period of such a candidate against the $\chi^2$ of the corresponding fit, coloured by planetary mass. All candidates have a mass below 45$M_E$, and 90\% of them have $m_c$ smaller than 20$M_E$, showing that only a very small mass companion could be "hidden" in the RV signal. Moreover, via the BIC, we calculated the $\chi^2$ value that needs to be reached to have (at least) weak evidence that a two-planet fit is better than a single-planet fit. The red dashed line denotes such a value. Our analysis did not retrieve the presence of any candidate body that passes the the BIC threshold, according to \cite{Kass1995} criterions.\\

\begin{figure}
    \centering
    \includegraphics[width=0.5\textwidth]{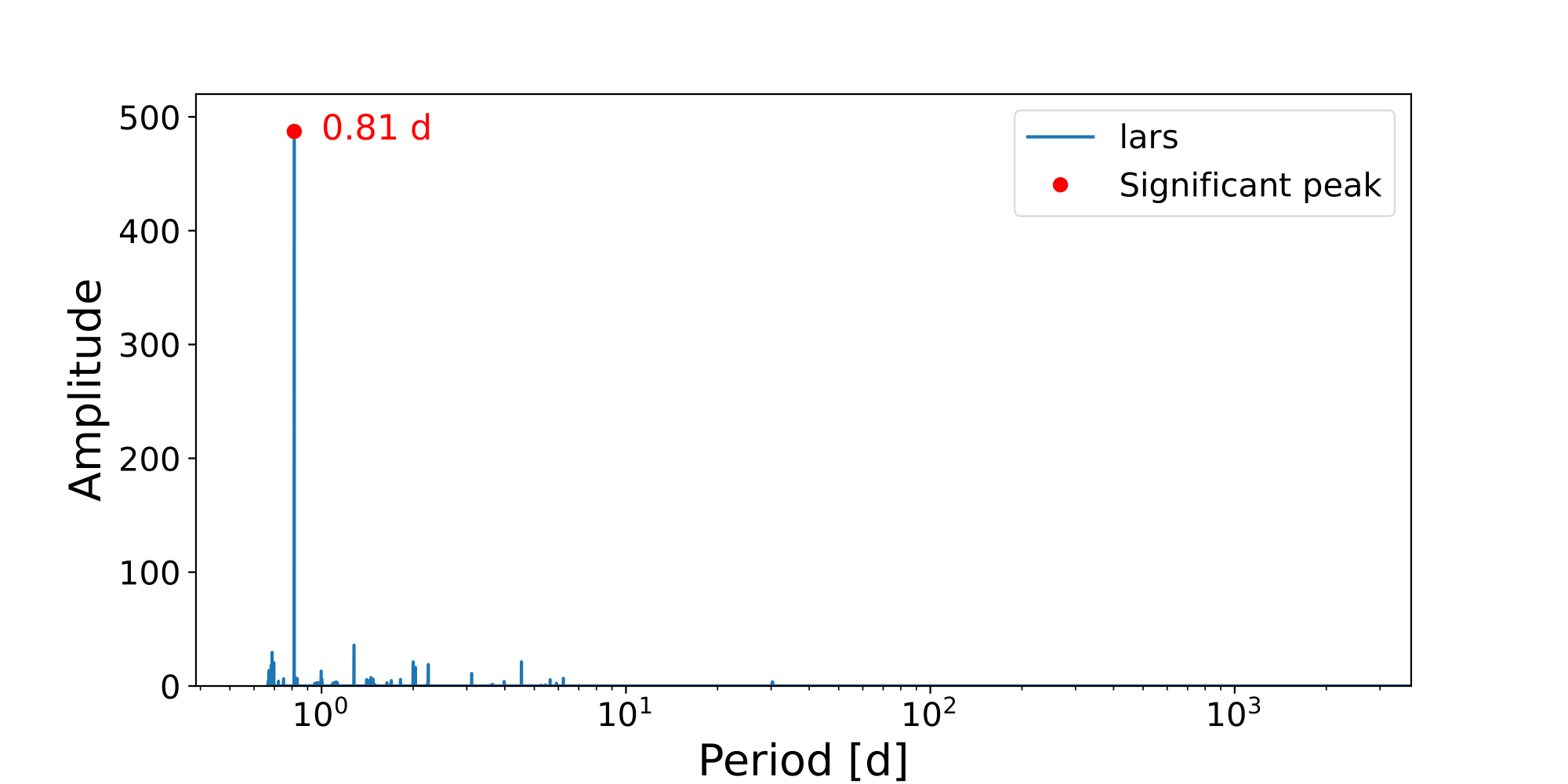}
    \includegraphics[width=0.5\textwidth]{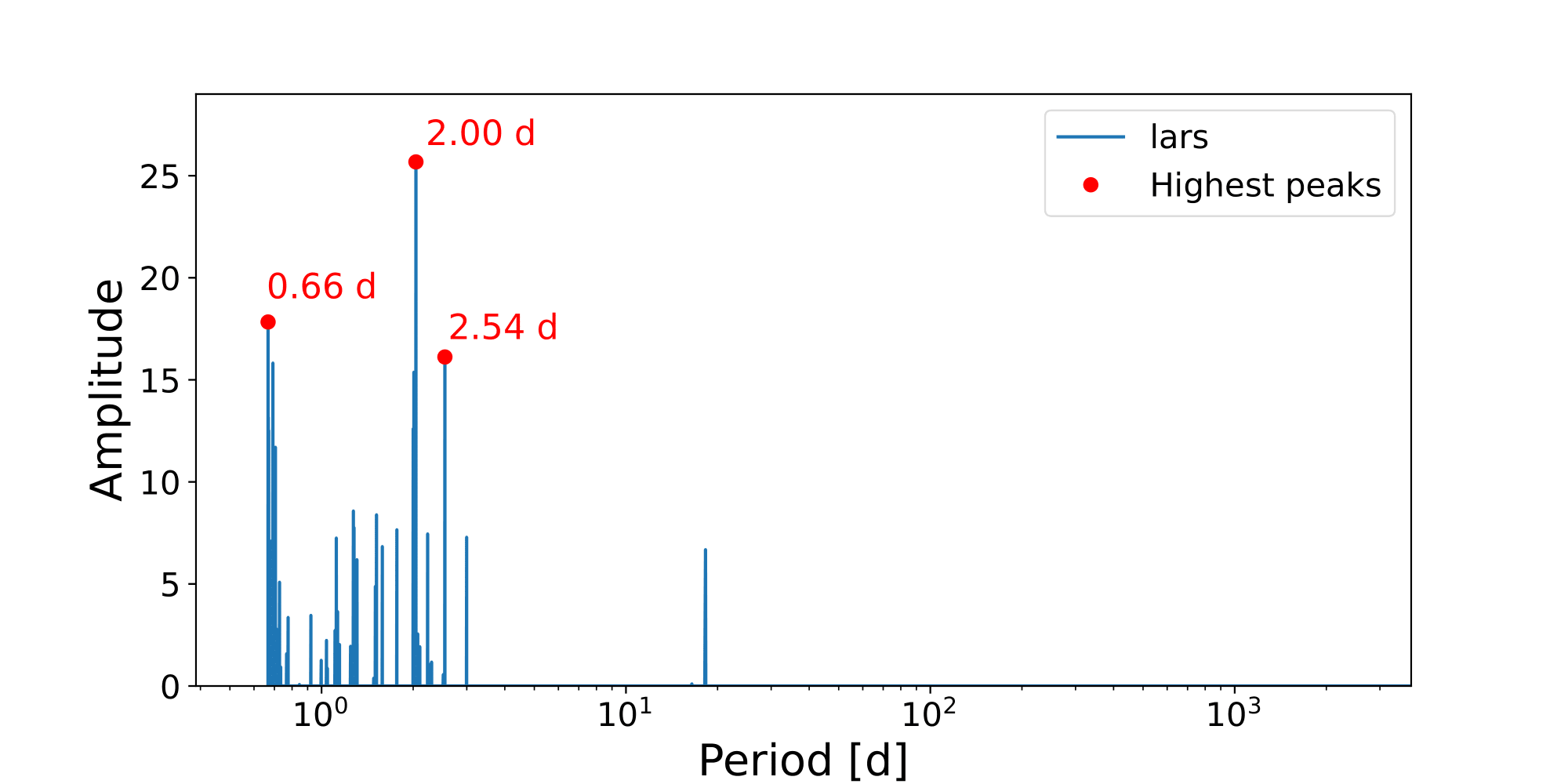}
    \caption{Periodograms of RVs.\\
    {\it Top plot} Periodogram of the RV dataset, showing the only significant peak at $\simeq$0.814 days, corresponding to the orbital period of WASP-43b.\\
    {\it Low plot} Periodogram of the residuals showing the three highest peaks, which do not pass the FAP test with a threshold at 1\%.}
    \label{fig:periodogram}
\end{figure}

\noindent In order to exclude any putative planetary companion in the system, we also used the $l_1$ periodogram analysis (\citealp{Hara2017}) which is similar to the Lomb-Scargle periodogram and was specifically designed to search for periodicities in radial velocities. We applied it on the whole RV dataset, separately on the single datasets and on the residuals after subtracting the RV signal produced by planet {\it b}, shown in the top and bottom plot of Figure~\ref{fig:periodogram}, respectively. In both cases, the in-transit points are excluded from the analysis. In the first case, the only significant peak with a False Alarm Probability (FAP) smaller than 1\% is the one at 0.81 days, the orbital period of WASP-43b. The periodogram of the residuals does not show any significant peak passing the FAP test, with a threshold at 1\%, meaning that only 1\% of periodograms calculated on random noise would have a peak amplitude equal to or greater than that observed in the periodogram. The peaks relate most probably to an artefact in the data, instead of a true period (see also the different y-amplitude between the two plots). \\ 

\begin{figure}
    \centering
    \includegraphics[width=0.5\textwidth]{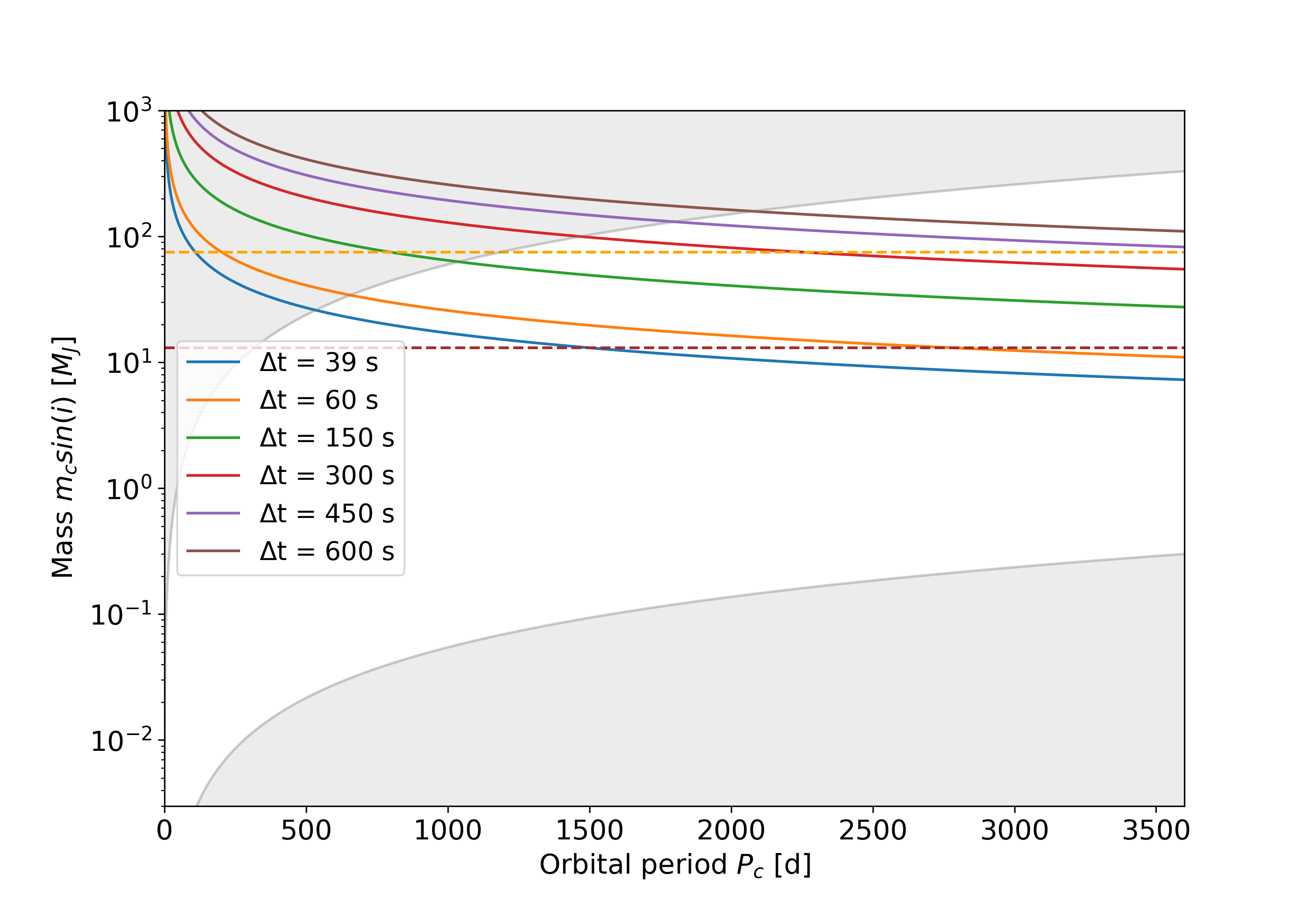}
    \caption{Results of the sensitivity analysis based on the light travel effect. The plot shows the relationship between the orbital period and the planetary mass of a planet responsible for 10, 25, 50, 75 and 100\% of the TTVs. The lowest blue curve reflects the possibility that the periastron precession rate observed is caused solely by the R{\o}mer effect (see text for more details). The yellow and brown horizontal lines represent the minimum stellar mass ($\simeq$75 $M_J$) and the minimum brown dwarf mass ($\simeq$13 $M_J$), respectively. The grey shaded regions represent the excluded area by RV analysis. The upper one indicates an  hypothetical planet causing an RV amplitude > 125 m s$^{-1}$. The lower one determines the sensitivity threshold of 5 m s$^{-1}$ (see text for more details).\\ 
    }
    \label{fig:Romer}
\end{figure}

\noindent We also report a sensitivity analysis based on the R{\o}mer effect (or light time effect; \citealp{Irwin1952}). If TTVs are due to an outer companion, this  would cause the system’s centre of mass to be shifted. The observed mid-transit and occultation times are affected by the time difference that it takes the light to travel from the far side to the near side of the orbit with respect to the observer (see the case of WASP-4b in \citealp{Harre2023b}, where the apparent tidal decay is explained by the light travel effect). For circular orbits, the TTV amplitude $\Delta t$ due to such an effect is (\citealp{Schneider2005}):
\begin{equation}
    \Delta t = 2 \frac{m_p}{M_\star} \frac{a \sin i}{c}.
\end{equation}
Subtracting the TTV signal of planet {\it b}, the maximum TTV residual is $\simeq$600 s, which is set as $\Delta t$ in the Equation. With this assumption, we calculated the corresponding putative planetary mass by changing its orbital period of WASP-43b from the period of WASP-43b (since it is an outer companion) to the length of the transit dataset ($\simeq$3\,700 days). We also considered the possibility that only part of the TTV residuals is due to the R{\o}mer effect and accounted for scenarios where $\Delta t$=10, 25, 50 and 75\% of such a TTV amplitude. Additionally, we considered the hypothesis that the periastron precession rate we observe in Section \ref{sec:complete_model} could be caused by the presence of a planet. The amplitude of the TTVs caused by an apsidal motion of 0.1727 $^\circ d^{-1}$ results in $\simeq$ 39 s. Therefore we also assumed $\Delta t$=39 s and repeated the study (see the blue line in Figure \ref{fig:Romer}). Figure \ref{fig:Romer} illustrates the relationship between the orbital period and the planetary mass and planetary mass for a planet that could shift the system's centre of mass and generate such a TTV amplitude. The two horizontal lines in the figure mark the minimum brown dwarf and stellar mass. The upper shaded gray area in the plot indicates the region where a hypothetical planet with given period and mass would be detectable by our RV data. Specifically, this area indicates the parameter space where any such planet would produce a radial velocity signal that exceeds the residuals left after subtracting the known planet {\it b} ($\simeq$ 125 m s$^{-1}$). Consequently, if a planet with these parameters existed, it would have been likely detected. \\
\indent In determining the detection threshold of our study, we consider the typical error of the RV data (conservatively we take their median $\simeq$5 m s$^{-1}$). 
The lower grey shaded region in the plot delineates the parameter space where hypothetical planets would induce a radial velocity semi-amplitude smaller than such a threshold. A hypothetical planet could potentially reside in the lower shaded region and still be undetected, having a mass $m_c$ less than $\simeq$0.3 $M_J$. The exact limiting value of mass depends on its combination with the orbital period value, as seen in the Figure. \\

\begin{figure}
    \centering
    \includegraphics[width=0.5\textwidth]{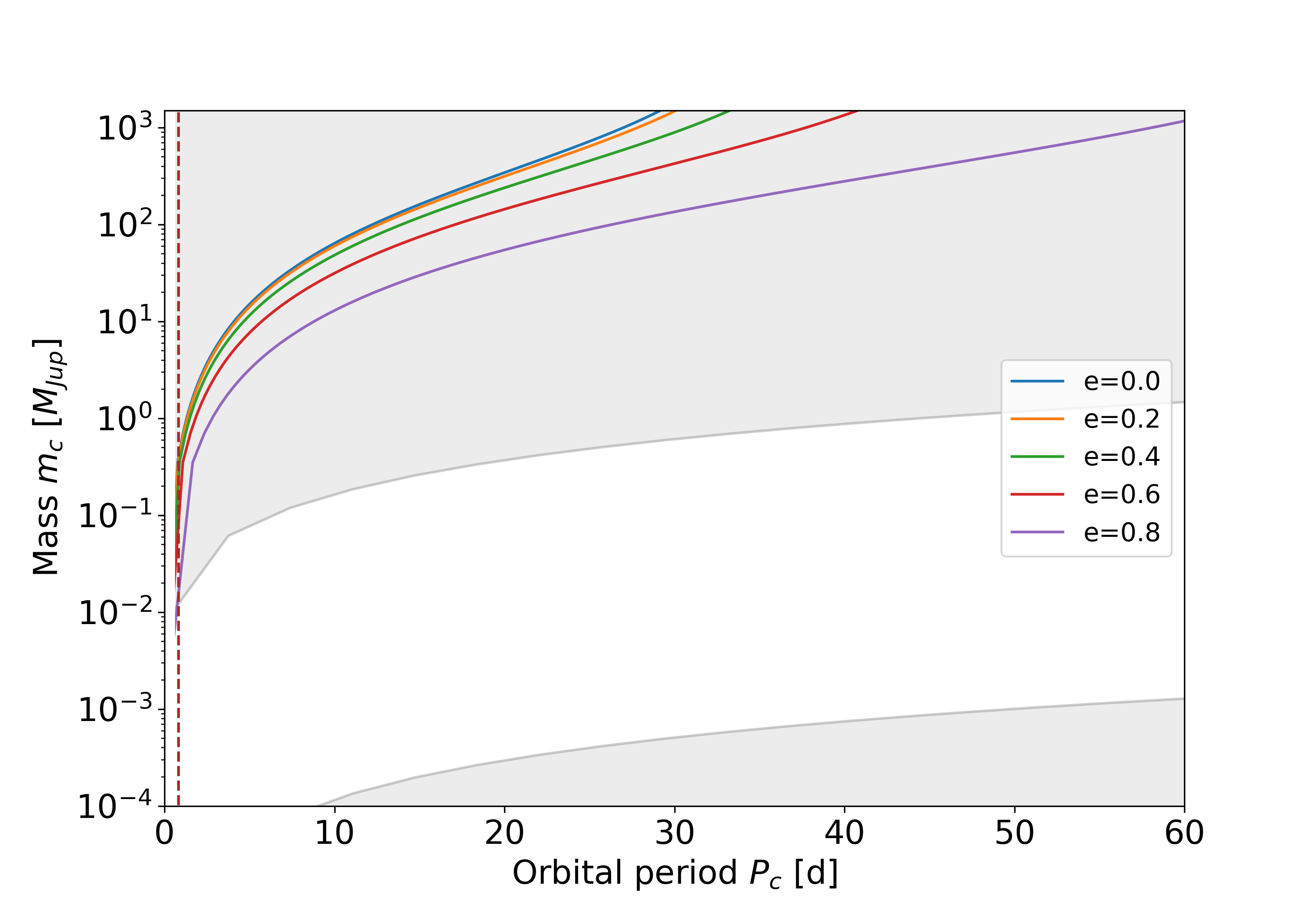}
    \caption{Sensitivity study on the mass and orbital period of a body causing the observed apsidal motion rate of $\simeq$ 0.1727 $^\circ$ d $^{-1}$, by varying its orbital eccentricity and mass, according of \cite{Borkovits2011}. The two grey shaded region represent the same as in Figure \ref{fig:Romer}, see text for more details. The vertical dashed line represents the orbital period of WASP-43b.}
    \label{fig:planetc_wdot}
\end{figure}

\noindent We also estimated the mass and period of an hypothetical third body contributing to the detected apsidal motion rate, using Equation 12 from \cite{Borkovits2011} in case of coplanar, eccentric, outer ($P_c$>$P_b$) orbit. Its amplitude is:
\begin{equation}
    \dot{\omega} \simeq \frac{3\pi}{2} \frac{m_c}{M_\star + m_b + m_c} \frac{P_b}{P_c^2} (1-e_c^2)^{-3/2}
\end{equation}
where $m_b$ and $m_c$ denote the masses of planet {\it b} and the putative companion {\it c}, $P_b$ and $P_c$ their orbital periods and $e_c$ the orbital eccentricity of body {\it c}. We subtracted the General Relativity contribution from the detected apsidal motion rate and calculated the mass and orbital period of a planet that would account the measured $\dot{\omega}$. The results of this analysis are shown in Figure \ref{fig:planetc_wdot}, where we vary the eccentricity of the orbit of the putative companion from 0 to 0.8. The two grey shaded area represent the same sensitivity limits as in Figure \ref{fig:Romer}. Since the upper shaded region encompasses all the plotted period-mass curves, we can exclude the possibility that the observed apsidal motion rate is caused by an unseen second body within the sensisivity limits expressed in the previous paragraph and Figure \ref{fig:Romer}.\\ 

\noindent In this section, we excluded the possibility of the presence of a companion in the system which may cause the detected high apsidal motion and orbital decay rate. The sensitivity of our analysis depends on the planetary orbital period. It goes down to a fraction of Jupiter's mass at period up to $\simeq$3\,700 days, the length of the time window of our dataset. Moreover, in Sections \ref{sec:longterm} and \ref{sec:complete_model}, we retrieved a long-term trend in RVs (or long-term acceleration, $\dot{V}_\gamma$) which is compatible with a non-accelerating system (see Table \ref{tab:model_parameters}), confirming that there is no hint of a second stellar or planetary body in our dataset.

\subsection{Stellar inclination}
\label{sec:RM}
\begin{figure}
    \centering
    \includegraphics[width=0.5\textwidth]{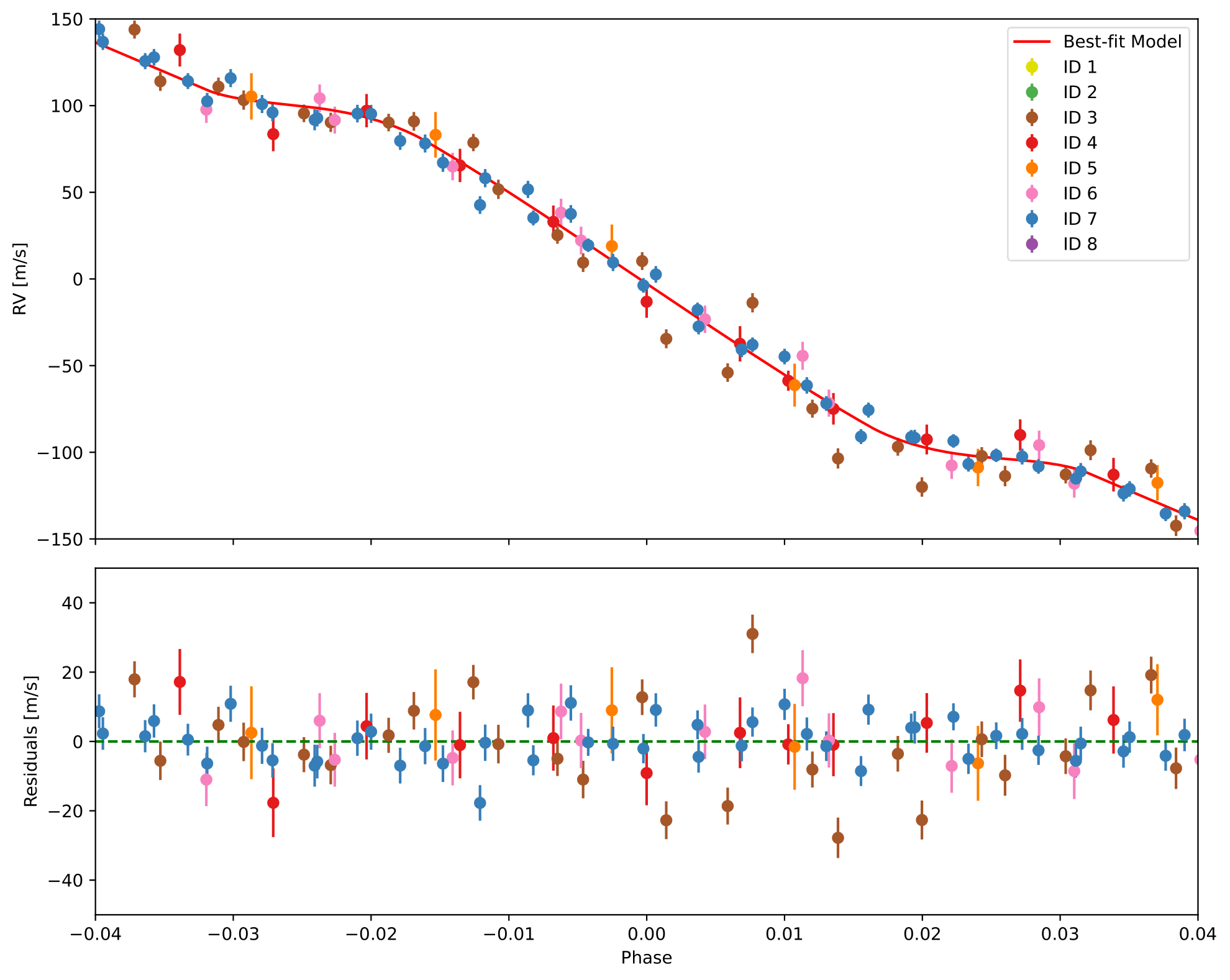}
    \caption{Best fitting model of the Rossiter-McLaughlin effect. Only datapoints within phase -0.04 and +0.04 are plotted to zoom into the transit region.}
    \label{fig:RossMcL}
\end{figure}

To derive the equations describing the periastron precession (see Appendix A in \citealp{Bernabo2024}) we assumed that the stellar axis is not (or slightly) inclined with respect to the normal to the orbital plane. If this is not the case, additional terms at higher orders can contribute to $\dot{\omega}$ (see \citealp{Kopal1959,Kopal1978}). The contribution of the stellar obliquity term would be an added or subtracted $\simeq$10\% of the total observed value.\\
\indent To test such a scenario, we modelled the (Holt-)Rossiter-McLaughlin (RM) effect (\citealp{Holt1893,Rossiter1924,McLaughlin1924}) in the system to derive the spin-orbit angle of WASP-43. We fitted the whole RV dataset in Table \ref{Tab:RV_data} - both in-transit and out-of-transit data - with a \texttt{C++} script that carries out the analytic calculation for the RM effect by \cite{Hirano2011} as well as the standard RV curve from the planet's Keplerian orbit. The fitted parameters are the RV semi-amplitude K, $\sqrt{e} \sin \omega$, $\sqrt{e} \cos \omega$, the stellar rotational velocity $v \sin i$, the mid-transit time, the sidereal orbital period, the impact parameter, the radius ratio $r_p/R_\star$, the scaled orbital distance $a/R_\star$, the projected stellar obliquity angle $\lambda$ and the RV offsets. The best-fitting model and the residuals are shown in Figure \ref{fig:RossMcL}. In Table \ref{tab:RossMcL} we report the results on the fitting parameters. We measured the projected stellar obliquity to be $\lambda$=2.03$\pm$1.32$^\degree$, confirming the results of \cite{Esposito2017} of a well-aligned star.

\begin{table*}[]
    \centering
    \caption{Results of the Rossiter-McLaughlin fit.}
    \begin{tabular}{l c c}
    \hline \hline
         Parameter & Symbol [Units] & Value\\
         \hline
         RV semi-amplitude & $K$ [km s$^{-1}$] & 0.5553 $\pm$ 0.0010 \\
         & $\sqrt{e} \cdot \sin\omega$ & -0.044 $\pm$ 0.016 \\
         & $\sqrt{e} \cdot \cos\omega$ & -0.024 $\pm$ 0.021 \\
         Stellar rotation & $v \sin i$ [km s$^{-1}$] & 2.14 $\pm$ 0.13  \\
         Stellar inclination & $\lambda$ [$^\circ$] & 2.03 $\pm$ 1.32 \\
         Scaled semi-major axis & $a/R_\star$ & 4.769 $\pm$ 0.029 \\
         Impact parameter & $b$ & 0.6839 $\pm$ 0.0070 \\
         Radius ratio & $r_p/R_\star$ & 0.1599 $\pm$ 0.0040 \\
         Transit mid-time & $T_{tr}$ [$\BJDTDB$-2~454~000] & 8912.107290 $\pm$ 0.000030 \\
         Sidereal period & $P_s$ [d] & 0.81347412 $\pm$ 0.00000006\\
    \end{tabular}
    \label{tab:RossMcL}
\end{table*}

\section{Conclusions}
In this paper, we applied the revised tidal theory of \cite{Csizmadia2019} and~\cite{Bernabo2024} on a third case study, the exoplanetary system WASP-43. We acquired new RV data with HARPS and merged this new dataset with the literature and three previously unpublished RV datasets. Together with the RVs, we fitted literature mid-transit and mid-occultation times, including the newly acquired JWST phase-curve and the unpublished transits in TESS Sector 62. \\
\indent First of all, we investigated a simple circular and eccentric one-planet model, finding statistical evidence of a non-zero eccentricity in the orbit. We also investigated three separated scenarios of tidal decay, periastron precession and long-term acceleration. Since other effects can mimic the consequences of tidal interaction, we ruled out the presence of a perturbing body, such as an extra planet, up to a period of $\simeq$3\,700 days and down to $\simeq$0.3$M_J$ and of long-term acceleration in WASP-43. \\
\indent Finally, we included all aforementioned effects in our model - long-term acceleration, periastron precession and tidal decay, as a result of the tidal interaction between the planet and the host star. For the first time in an exoplanetary system we detect simultaneously two outcomes of tidal interactions: tidal decay and apsidal motion. We confirm that the orbit of the system is decaying at a rate of $\dot{P}_a$=(-1.99 $\pm$ 0.50) ms yr$^{-1}$ and the first time in the system, we detected apsidal motion at a rate of $\dot{\omega}$= (0.1727$^{+0.0083}_{-0.0089}$)$^\circ$ d$^{-1}$ = (621.72 $^{+29.88}_{-32.04}$)$^{\prime\prime}$d$^{-1}$.\\
\indent We analysed a broad range of scenarios to explain such a high apsidal motion rate. Assuming a planetary rotational period synchronous to the orbital period - as predicted by the timescale of synchronization due to tidal interactions - the value of second-order fluid Love number we derive is above the physical limit of 1.5. We tried to explain the measured $\dot{\omega}$ by relaxing the assumption of synchronization of the planetary rotational period to the orbital period, but this is not supported by the recent JWST findings on the high day-to-night side temperature difference. This scenario could be further investigated in the future with e.g. measurements on the Rossiter-McLaughlin effect on the planet and line broadening detections. We also test the possibility of a contribution to $\dot{\omega}$ from the stellar inclined spin-orbit axis, but the RM analysis confirms that the projected obliquity plane is well-aligned. We also discard the presence of a perturber in the system from Gaia DR3 observations, a sensitivity analysis, the study of the $l_1$ periodogram and from fitting the RV dataset with an extra putative planet. Gaia Data Release 4 may confirm or change the picture by revealing a companion via astrometry. \\
\indent In conclusion, we have considered various astrophysical scenarios, but most of them encounter difficulties in explaining the results we obtain. We leave the question open for further investigation, highlighting that in research, the questions raised are often more significant than the answers found, driving further inquiry and deeper understanding.


\vspace{1cm}

Tables A.1, A.2 and B.1 are available in electronic form at the CDS via anonymous ftp to cdsarc.u-strasbg.fr (130.79.128.5) or via http://cdsweb.u-strasbg.fr/cgi-bin/qcat?J/A+A/. 

\begin{acknowledgements}
\indent We acknowledge the support of DFG Research Unit 2440: "Matter Under Planetary Interior Conditions: High Pressure, Planetary, and Plasma Physics" and of DFG grants RA 714/14-1 within the DFG Schwerpunkt SPP 1992, Exploring the Diversity of Extrasolar Planets. \\
\indent A.M.S.S. and J.-V.H. acknowledge support from the Deutsche Forschungsgemeinschaft grant SM 486/2-1 within the Schwerpunktprogramm SPP 1992 "Exploring the Diversity of Extrasolar Planets". \\
\indent Project no. C1746651 has been implemented with the support provided by the Ministry of Culture and Innovation of Hungary from the National Research, Development and Innovation Fund, financed under the NVKDP-2021 funding scheme.\\
\indent The work is based on observations made with the HARPS instrument on the ESO 3.6-m telescope at La Silla Observatory under programme IDs 089.C-0151, 096.C-0331, 0102.C-0820, 0104.C-0849.\\
\indent This research used the facilities of the Italian Center for Astronomical Archive (IA2) operated by INAF at the Astronomical Observatory of Trieste.\\
\indent This research has made use of the NASA Exoplanet Archive, which is operated by the California Institute of Technology, under contract with the National Aeronautics and Space Administration under the Exoplanet Exploration Program.\\
\indent L. M. B. would like to thank M. Murphy for reanalysing for us Spitzer Space Telescope lightcurves without priors and those observers who provided unpublished transits of WASP-43b: M. Mifsud, I. Peretto and S. Lora, J.-P. Vignes, S. Foschino, A. Popowicz, M. Szkudlarek, and especially C. Knight, who kindly conducted the observations at his observatory. L. M. B. also acknowledges A. Triaud for the use of the public data used in this work (RV data ID 3) and Nicolas Iro for the interesting discussion on the rotation of WASP-43b.\\
\indent The authors gratefully acknowledge the scientific support and HPC resources provided by the German Aerospace Center (DLR). The HPC system CARA is partially funded by "Saxon State Ministry for Economic Affairs, Labour and Transport" and "Federal Ministry for Economic Affairs and Climate Action".

\end{acknowledgements}



\nocite{*}
\bibliographystyle{aa} 
\bibliography{aa.bib} 

\newpage
\onecolumn
\begin{appendix}
\section{Transit and occultation mid-timings}
\label{app:TROCCdata}
Tables of transit and occultation mid-times in the literature and used in this work, along with details on the observations.
\setcounter{table}{0}
\renewcommand\thetable{A.\arabic{table}} 

\begin{table*}[h]
\caption{Mid-transit times.}
\resizebox{\textwidth}{!}{
    \centering
    \begin{tabular}{l l l l l}
    \hline
    \hline
    {\bf {\large Cycle number}} & 
    {\bf {\large T$_\mathrm{mid}$ }}[days] & 
    {\bf {\large Uncertainty}} [days] & Observations & {\bf {\large Reference}}\\
N & $\BJDTDB$ - 2\,450\,000 & & & \\
\hline
0 & 5528.86823 & $\pm$0.00014 & WASP-South, SuperWASPNorth, TRAPPIST, EulerCAM & \cite{Hellier2011} \\
11 & 5537.81648 & $^{+0.00048} _{-0.00043}$ & TRAPPIST-South, ESO La Silla Observatory, Chile & \cite{Gillon2012} \\
22 & 5546.76490 & $\pm$0.00020 & TRAPPIST-South, ESO La Silla Observatory, Chile & \cite{Gillon2012} \\
27 & 5550.83228 & $^{+0.00014}_{-0.00013}$ & TRAPPIST-South, ESO La Silla Observatory, Chile & \cite{Gillon2012} \\
38 & 5559.78038 & $\pm$0.00021 & TRAPPIST-South, ESO La Silla Observatory, Chile & \cite{Gillon2012}\\
... & ... & ... & ... & ...\\
\end{tabular}
  }
 \tablefoot{We list cycle number referred to~\cite{Hellier2011}, mid-transit point in $\mathrm{BJD}_\mathrm{TDB}$, its errorbar, details on the observations and reference or observer. \\
 Only the first rows of the Table are shown. The complete Table is available in electronic form.
}
    \label{Tab:transits1}
\end{table*}

\begin{table*}[h]
\centering
\caption{Mid-occultation times.}
\resizebox{\textwidth}{!}{%
    \centering
    
    \begin{tabular}{l l l l l}
    \hline
    \hline
    {\bf {\large Cycle number}} &{\bf {\large T$_\mathrm{mid}$ }}[days] & {\bf  {\large Uncertainty}} [days] & Observations & {\bf {\large Reference }}\\
N & $\BJDTDB$ - 2\,450\,000 & & & \\
\hline
13 & 5539.8519 & $^{+0.0016}_{-0.0015}$ & VLT/HAWK-I, ESO Paranal Observatory, Chile & \cite{Blazek2022} from \cite{Gillon2012} \\
299 & 5772.5043 & $\pm$0.0003 & Spitzer Space Telescope & \cite{Blecic2014} \\
300 & 5773.3177 & $\pm$0.0003 & Spitzer Space Telescope & \cite{Blecic2014} \\
565 & 5988.8844 & $\pm$0.0019 & Canada–France–Hawaii Telescope & \cite{Wang2013} \\
613 & 6027.9364 & $\pm$0.0018 & Canada–France–Hawaii Telescope & \cite{Wang2013} \\
... & ... & ... & ... & ...\\
\end{tabular}
    }
    \tablefoot{We list cycle number, referred to~\cite{Hellier2011}, mid-transit point in $\mathrm{BJD}_\mathrm{TDB}$, its errorbar, details on the observations and reference or observer. Light travel time across the system has been subtracted from $T_\mathrm{mid}$.\\
    Only the first rows of the Table are shown. The complete Table is available in electronic form.}
    \label{Tab:occultations}
\end{table*}

\section{RV datasets}
\label{app:RVdata}
Table of archive and unpublished RV datasets used in this work.

\setcounter{table}{0}

\renewcommand\thetable{B.\arabic{table}} 
\begin{table}[h]
    \centering
    \caption{Radial Velocity datasets used in this work.}
    \begin{tabular}{c c c c}
\hline
\hline
Instr. ID & $\mathrm{BJD}_\mathrm{TDB}$ & $RV_\mathrm{obs}$ & Uncertainty  \\
& $-2~450~000$ & [km s$^{-1}$] & [km s$^{-1}$] \\
\hline
1 & 5205.7602 & –3.058 & 0.013 \\
1 & 5325.6240 & –4.041 & 0.021 \\
1 & 5327.5753 & –3.430 & 0.026 \\
1 & 5328.5447 & –3.067 & 0.014 \\
1 & 5334.5038 & –3.821 & 0.018 \\
... & ... & ... & ...\\
\hline
\end{tabular}
 \tablefoot{The ID number corresponds to the description in Section~\ref{subsec:RVs}.\\
 Only the first rows of the Table are shown. The complete Table is available in electronic form.}
 \label{tab:RVs1}
\end{table}

\end{appendix}

\end{document}